\PassOptionsToPackage{unicode}{hyperref}
\PassOptionsToPackage{hyphens}{url}
\PassOptionsToPackage{dvipsnames,svgnames,x11names}{xcolor}
\documentclass[11pt,a4paper]{article}
\usepackage[margin=2.5cm,headheight=14pt]{geometry}
\usepackage{xcolor}
\usepackage{amsmath,amssymb}
\usepackage[T1]{fontenc}
\usepackage[utf8]{inputenc}
\usepackage{textcomp} 
\usepackage{mathptmx} 
\DeclareUnicodeCharacter{2212}{\ensuremath{-}}
\IfFileExists{upquote.sty}{\usepackage{upquote}}{}
\IfFileExists{microtype.sty}{
  \usepackage[]{microtype}
  \UseMicrotypeSet[protrusion]{basicmath} 
}{}
\IfFileExists{parskip.sty}{%
  \usepackage{parskip}
}{
  \setlength{\parindent}{0pt}
  \setlength{\parskip}{6pt plus 2pt minus 1pt}}
\makeatletter
\ifx\paragraph\undefined\else
  \let\oldparagraph\paragraph
  \renewcommand{\paragraph}{
    \@ifstar
      \xxxParagraphStar
      \xxxParagraphNoStar
  }
  \newcommand{\xxxParagraphStar}[1]{\oldparagraph*{#1}\mbox{}}
  \newcommand{\xxxParagraphNoStar}[1]{\oldparagraph{#1}\mbox{}}
\fi
\ifx\subparagraph\undefined\else
  \let\oldsubparagraph\subparagraph
  \renewcommand{\subparagraph}{
    \@ifstar
      \xxxSubParagraphStar
      \xxxSubParagraphNoStar
  }
  \newcommand{\xxxSubParagraphStar}[1]{\oldsubparagraph*{#1}\mbox{}}
  \newcommand{\xxxSubParagraphNoStar}[1]{\oldsubparagraph{#1}\mbox{}}
\fi
\makeatother

\usepackage{longtable,booktabs,array}
\usepackage{calc} 
\usepackage{etoolbox}
\makeatletter
\patchcmd\longtable{\par}{\if@noskipsec\mbox{}\fi\par}{}{}
\makeatother
\IfFileExists{footnotehyper.sty}{\usepackage{footnotehyper}}{\usepackage{footnote}}
\makesavenoteenv{longtable}
\usepackage{graphicx}
\makeatletter
\newsavebox\pandoc@box
\newcommand*\pandocbounded[1]{
  \sbox\pandoc@box{#1}%
  \Gscale@div\@tempa{\textheight}{\dimexpr\ht\pandoc@box+\dp\pandoc@box\relax}%
  \Gscale@div\@tempb{\linewidth}{\wd\pandoc@box}%
  \ifdim\@tempb\p@<\@tempa\p@\let\@tempa\@tempb\fi
  \ifdim\@tempa\p@<\p@\scalebox{\@tempa}{\usebox\pandoc@box}%
  \else\usebox{\pandoc@box}%
  \fi%
}
\def\fps@figure{htbp}
\makeatother

\usepackage{fancyhdr}
\usepackage{needspace}
\fancypagestyle{plain}{%
  \fancyhf{}%
  \fancyhead[R]{\thepage}%
  }
\usepackage{bookmark}
\IfFileExists{xurl.sty}{\usepackage{xurl}}{} 
\hypersetup{
  pdftitle={Large language models underestimate and partly misrepresent cultural variation in everyday norms},
  colorlinks=true,
  linkcolor={blue},
  filecolor={Maroon},
  citecolor={Blue},
  urlcolor={Blue},
  pdfcreator={LaTeX via pandoc}}

\title{Large language models underestimate and partly misrepresent
cultural variation in everyday norms}
\author{}
\date{}
\begin{document}
\maketitle

Kimmo Eriksson\textsuperscript{1,2}, Irina
Vartanova\textsuperscript{1,3}, Pontus Strimling\textsuperscript{1,4}

\textsuperscript{1}Institute for Futures Studies, Stockholm, Sweden\\
\textsuperscript{2}Department of Business and Mathematics, Mälardalen
University, Västerås, Sweden\\
\textsuperscript{3}Department of Women's and Children's Health, Uppsala
University, Uppsala, Sweden\\
\textsuperscript{4}Institute for Analytical Sociology, Linköping
University, Norrköping, Sweden

Correspondence: Kimmo Eriksson, Department of Business and Mathematics,
Mälardalen University, Box 883, 721 23 Västerås, Sweden. Email:
kimmo.eriksson@mdu.se

Classification: Social and Political Sciences; Psychological and
Cognitive Sciences


\subsection{Significance Statement}\label{significance-statement}

AI assistants need to anticipate how people judge everyday behavior, a
task complicated by norms varying across cultures. This study indicates
that frontier large language models (LLMs) do not yet provide reliable
knowledge of cultural differences in norms. We reached this conclusion
by comparing estimates from four LLMs with survey ratings of 150
everyday scenarios in 90 societies. The LLMs severely underestimated the
magnitude of cultural differences and, for many scenarios, only weakly
identified which societies judged the behavior more or less acceptable.
Accuracy was not uniformly poor: cultural differences were represented
more accurately for certain behaviors, especially kissing and flirting;
errors were somewhat smaller for norms in more developed societies; and
prompting in local survey languages produced modest improvements.

\subsection{Abstract}\label{abstract}

A key aspect of culture is a society's norms about everyday behavior.
How accurately do large language models (LLMs) represent cultural
differences in such norms? To answer this question we used the recent
Global Study of Everyday Norms (GSEN), which collected ratings of 150
scenarios in 90 societies, as the human benchmark. We prompted GPT-5 to
estimate each society's average rating for every scenario, and later
repeated the benchmark in three other LLMs: GPT-5.4, Claude Opus 4.6,
and Gemini 3.1 Pro. Compared to GSEN estimates, all four LLMs
misrepresented cultural variation in two ways. First, they greatly
underestimated its magnitude, estimating differences between societies
to be, on average, less than half their measured size. Second, for many
scenarios the LLMs poorly identified the pattern of variation, that is,
which societies judged the behavior less acceptable and which societies
judged it more acceptable. The pattern of variation was identified
better for scenarios that elicit concerns about vulgarity, especially
scenarios involving kissing and flirting. We also found that norms in
more developed societies tended to be estimated somewhat more
accurately, and that prompting in local survey languages rather than
English produced only a modest improvement in accuracy. Local-language
prompting also reduced, but did not remove, the underestimation of
between-society differences. Cultural differences in everyday norms are
only weakly and unevenly represented by LLMs.

\subsection{Introduction}\label{introduction}

As AI assistants and social robots enter everyday settings, they need to
anticipate how people will judge behavior. Social norms govern
innumerable judgments of this kind, and each judgment depends on the
situation: laughing is appropriate at a party and out of place at a
funeral. The rules are largely unwritten, and a system that misjudges
them can disrupt the interaction it was built to support. What large
language models (LLMs) know about social norms, and where that knowledge
fails, therefore matters for their safe use (Eriksson, Karlsson, et al.,
2026). Evidence from a single society is favorable: a study of everyday
norms in the United States found that LLMs estimate them more accurately
than most human participants do. Here, we ask a different question:
whether LLMs accurately represent variation in everyday norms across
societies.

Norms are an integral part of culture (Gelfand et al., 2011). The same
behavior can be acceptable in one society and condemned in another.
Previous studies of LLMs have assessed related forms of cultural
knowledge. Tao et al.~(2024) compared country-conditioned LLM responses
with nationally representative data on broad cultural values, while Zhao
et al.~(2024) tested whether LLMs could predict responses to value
questions from respondents' cultural and demographic characteristics.
Ramezani and Xu (2023) examined whether English-language models captured
differences between countries on contested moral issues, such as divorce
and homosexuality. Moving closer to everyday behavior, Rao et al.~(2025)
tested whether LLMs could judge the social acceptability of
etiquette-related situations drawn from 75 countries. Together, these
studies show that LLM performance varies with cultural context and that
prompting models with a country or cultural identity does not ensure
accurate cultural alignment.

Everyday norms present a distinct empirical problem. Societies do not
differ only in broad values or in their positions on a limited set of
contentious issues. They have norms for numerous specific behaviors in
particular situations, and the pattern of differences can change from
one behavior to another. Assessing whether an LLM represents this
variation therefore requires a common set of behaviors judged in many
societies. It also requires estimates of the degree to which each
society accepts each behavior in a given situation, rather than binary
classifications of scenarios as acceptable or unacceptable. The Global
Study of Everyday Norms (GSEN; Eriksson et al., 2025) provides this
structure: more than 25,000 participants in 90 societies rated 150
scenarios formed by placing 15 everyday behaviors in 10 situations.

Our primary question is how accurately LLMs represent cultural variation
in norms. We decompose it into two orthogonal questions: how accurately
LLMs estimate the magnitude of cultural variation and how accurately
they identify its pattern, that is, which societies find a given
scenario more or less acceptable. The magnitude could be estimated
incorrectly even if the pattern were captured accurately, or vice versa.

For a fuller understanding, we examine three additional questions.
First, is LLM accuracy consistent across everyday scenarios or does it
depend strongly on properties of the behavior being judged? LLMs might
represent cultural differences more accurately for behaviors that elicit
certain concerns than for others (Eriksson, Strimling, et al., 2026).

Second, how does the accuracy of LLM estimates vary across societies?
Poor representation of cultural variation could arise because LLMs
estimate norms accurately in some societies but poorly in others, with
previous research suggesting lower accuracy for less economically
developed or non-English-speaking populations (Tao et al., 2024).

Third, does accuracy improve when an LLM is prompted in the language in
which a society's survey was administered rather than in English? Prompt
language can affect the cultural values expressed by LLMs, although
existing evidence suggests that local-language prompting does not
necessarily overcome their cultural defaults (Bulté \& Rigouts Terryn,
2026). If local-language prompts elicit cultural knowledge that English
prompts do not, an English-language benchmark may underestimate how
accurately LLMs can represent cultural variation.

We addressed these questions in two stages. We first conducted an
exploratory benchmark using GPT-5 and then replicated the benchmark with
three later-released LLMs: GPT-5.4, Claude Opus 4.6, and Gemini 3.1 Pro.
Before collecting the replication data, we preregistered two findings
from the exploratory benchmark for confirmation in these three models;
see Results for details.

\subsection{Results}\label{results}

\subsubsection{How accurately do LLMs represent cultural variation in
norms?}\label{how-accurately-do-llms-represent-cultural-variation-in-norms}

\paragraph{LLMs make societies look too
similar}\label{llms-make-societies-look-too-similar}

In the exploratory GPT-5 analysis the estimated differences between
societies were much smaller than the differences the survey measured.
For each scenario, we compared the standard deviation (SD) of the LLM
estimates across societies with the SD of the survey means (Fig. 1).
Averaged over the 150 scenarios, the SD in GPT-5's estimates was less
than half (49\%) as large as the SD of the survey means. For the
replication study, we preregistered that the mean ratio of LLM to survey
between-society variation would have a 95\% bootstrap interval entirely
below one. This was supported for all three confirmatory LLMs (Table 1).
Across all four LLMs, estimated differences between societies were less
than half their measured size. Correcting the survey's between-society
SD for sampling noise produced similar estimates: variation in LLM
estimates was 47\% to 54\% of the adjusted survey variation. The
bootstrap interval also remained below one when behaviors rather than
scenarios were resampled (Table S1).

\pandocbounded{\includegraphics[keepaspectratio]{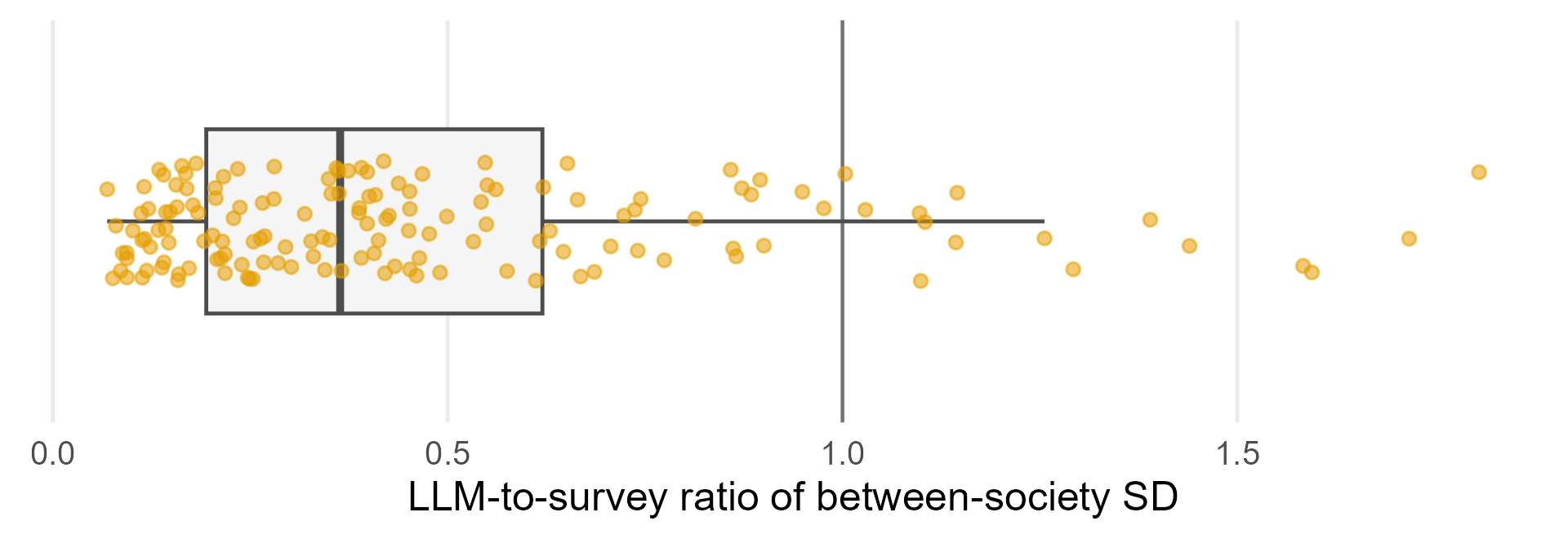}}

\textbf{Figure 1.} The four LLMs estimate smaller differences between
societies than the survey measures. For each scenario, the plotted value
is the ratio of the between-society SD of the LLM estimates to the
between-society SD of the survey means (averaged across the four LLMs;
see Fig. S1 for results per LLM). A ratio of 1 means equal variation,
values below 1 mean the LLMs underestimate between-society variation,
and values above 1 mean they overestimate it. The box marks the median
and the quartiles, and the overlaid dots mark the individual ratios.

\needspace{8\baselineskip}
\textbf{Table 1.} Cross-cultural benchmark for the four LLMs on the GSEN
90 × 150 grid.

\begin{longtable}[]{@{}
  >{\raggedright\arraybackslash}p{(\linewidth - 10\tabcolsep) * \real{0.2339}}
  >{\raggedright\arraybackslash}p{(\linewidth - 10\tabcolsep) * \real{0.2329}}
  >{\raggedright\arraybackslash}p{(\linewidth - 10\tabcolsep) * \real{0.1092}}
  >{\raggedright\arraybackslash}p{(\linewidth - 10\tabcolsep) * \real{0.2329}}
  >{\raggedright\arraybackslash}p{(\linewidth - 10\tabcolsep) * \real{0.1097}}
  >{\raggedright\arraybackslash}p{(\linewidth - 10\tabcolsep) * \real{0.0813}}@{}}
\toprule\noalign{}
\begin{minipage}[b]{\linewidth}\raggedright
LLM
\end{minipage} & \begin{minipage}[b]{\linewidth}\raggedright
Compression
\end{minipage} & \begin{minipage}[b]{\linewidth}\raggedright
Cross-cultural r
\end{minipage} & \begin{minipage}[b]{\linewidth}\raggedright
Vulgarity partial r
\end{minipage} & \begin{minipage}[b]{\linewidth}\raggedright
Within-society MAE
\end{minipage} & \begin{minipage}[b]{\linewidth}\raggedright
Level offset
\end{minipage} \\
\midrule\noalign{}
\endhead
\bottomrule\noalign{}
\endlastfoot
GPT-5 & 0.49 {[}0.42, 0.57{]} & 0.27 & 0.52 {[}0.29, 0.63{]} & 0.87 &
-0.32 \\
GPT-5.4 & 0.43 {[}0.38, 0.49{]} & 0.24 & 0.51 {[}0.08, 0.65{]} & 0.93 &
-0.64 \\
Claude Opus 4.6 & 0.47 {[}0.41, 0.53{]} & 0.23 & 0.54 {[}0.29, 0.65{]} &
0.68 & -0.19 \\
Gemini 3.1 Pro & 0.47 {[}0.41, 0.55{]} & 0.33 & 0.48 {[}0.16, 0.62{]} &
0.85 & -0.44 \\
\end{longtable}

\emph{Note}. Compression is the mean ratio of LLM to survey
between-society SD across 150 scenarios. Cross-cultural \emph{r} is the
correlation between LLM estimates and survey means across societies,
calculated separately for each scenario and then averaged across
scenarios. Vulgarity partial \emph{r} is the partial correlation between
cross-cultural accuracy and the vulgarity marker, controlling for survey
between-society SD. Within-society MAE is the mean absolute difference
between an LLM estimate and the corresponding survey mean on the -2.5 to
2.5 scale, averaged across societies. Level offset is the corresponding
signed mean difference. Brackets give 95\% percentile bootstrap
intervals from 2,000 resamples, with scenarios resampled for compression
and whole behaviors resampled for vulgarity partial \emph{r}.

\paragraph{LLMs identify the pattern of cultural differences only
weakly}\label{llms-identify-the-pattern-of-cultural-differences-only-weakly}

To assess how well LLMs identify the pattern of cultural differences in
norms, we correlated LLM estimates with survey means across societies
separately for each scenario, then averaged these correlations over
scenarios. The means ranged from .23 to .33 across the four LLMs (Table
1); correcting for sampling noise in the survey means and for the
reliability of repeated LLM answers increased the range somewhat to .26
to .40 (Table S2). Between 49\% and 66\% of scenarios fell below
\emph{r} = .30, depending on the LLM. However, the correlation was
considerably higher for certain scenarios (Fig. S2). We next examine
whether this variation in accuracy was systematically related to the
behavior being judged.

\subsubsection{Does accuracy depend on the
behavior?}\label{does-accuracy-depend-on-the-behavior}

\paragraph{LLMs represent cultural differences better for some behaviors
than for
others}\label{llms-represent-cultural-differences-better-for-some-behaviors-than-for-others}

Cross-cultural accuracy varied substantially across scenarios. To
examine whether that variation was structured by behavior, we averaged
the scenario-level correlations across the ten situations in which each
of the 15 behaviors appears. The resulting behavior-level accuracies
varied widely but were highly consistent across the four LLMs. At one
end, the LLMs distinguished societies relatively accurately for
behaviors such as kissing and flirting. At the other, they distinguished
societies poorly for ordinary behaviors such as talking and reading a
newspaper. Figure 2 shows this behavior-level ordering, averaged across
the four LLMs; the supplement gives the per-model breakdown (Fig. S3).

\pandocbounded{\includegraphics[keepaspectratio]{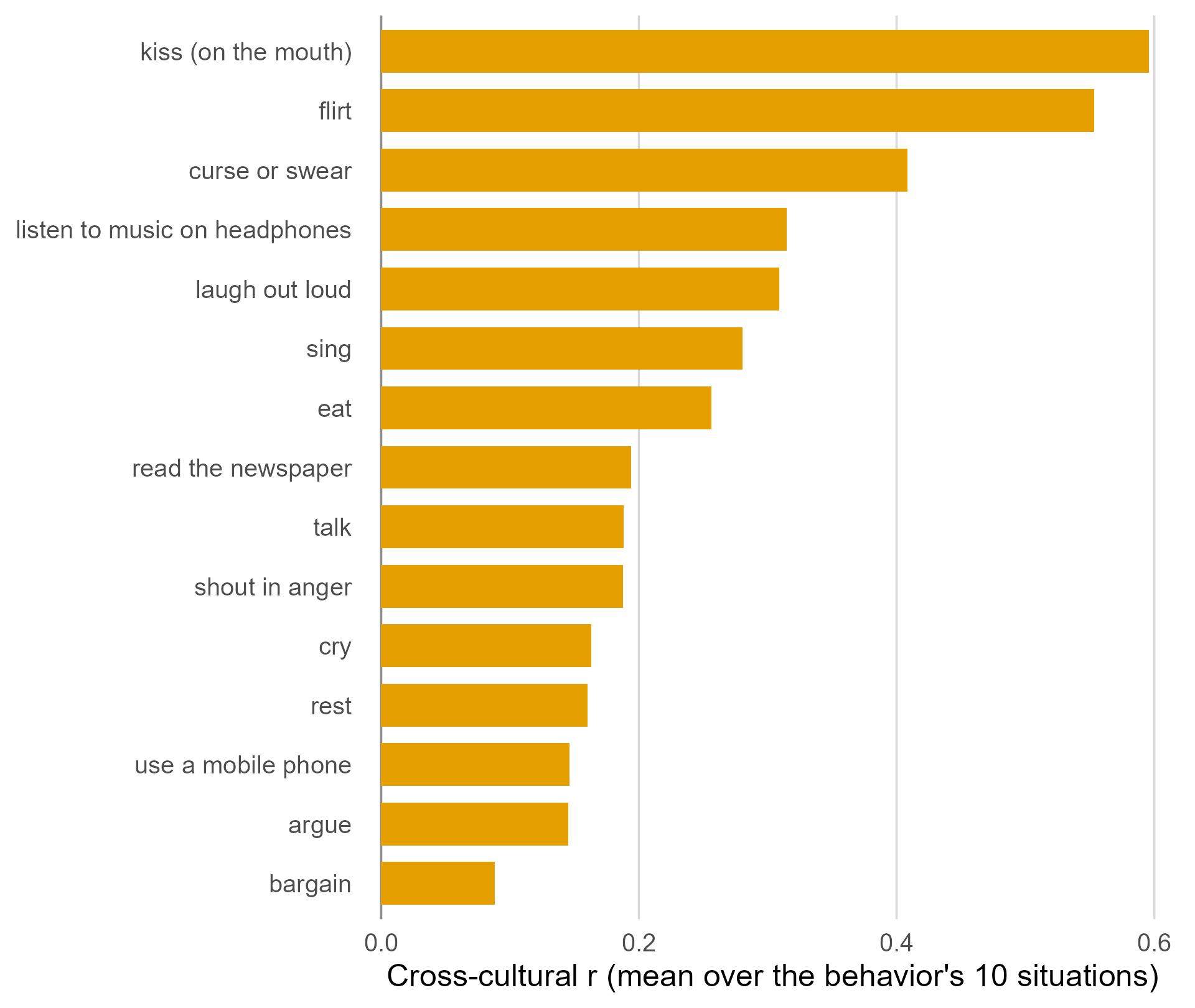}}

\textbf{Figure 2.} Cross-cultural accuracy by behavior, averaged across
the four LLMs. Each bar is a two-step average: for each LLM, the mean
cross-cultural correlation across the 10 situations in which the
behavior appears, then averaged across the four LLMs. Behaviors are
ordered by that same across-LLM mean, and the four LLMs agree closely
about the ordering (pairwise \emph{r} = .79 to .90).

What distinguishes the behaviors for which the LLMs represented cultural
differences relatively well? The behaviors near the top of the ordering
were not simply those judged most inappropriate or those on which
societies differed most. Instead, several involved conduct that
respondents often identified as vulgar. The GSEN asked respondents to
identify the main problem, if any, raised by each scenario. We defined
the vulgarity marker as the proportion of concern responses that
identified vulgarity rather than inconsiderateness or lacking sense.
Behaviors involving kissing and flirting received high vulgarity scores,
whereas behaviors such as talking received low scores (Table S3).

In the exploratory GPT-5 analysis, the vulgarity marker predicted
cross-cultural accuracy after controlling for the amount of
between-society variation available to predict (Fig. S4). We
preregistered this association for confirmation in the three additional
LLMs. The partial correlations were .51 for GPT-5.4, .54 for Claude Opus
4.6, and .48 for Gemini 3.1 Pro (Table 1). All three associations were
positive and met the preregistered decision rule of HC3 \emph{p}
\textless{} .05; in each case, \emph{p} \textless{} .001. The
association was therefore confirmed in all three LLMs. At the behavior
level shown in Figure 2, accuracy and the vulgarity marker correlated at
\emph{r} = .88 when accuracy was averaged across the four LLMs; the
corresponding correlations for the individual LLMs ranged from .79 to
.88. We next examined whether the association was specific to vulgarity
or extended to the other measured concerns. Vulgarity predicted
cross-cultural accuracy more strongly than either inconsiderateness or
lacking sense. With each concern measured as its proportion of all five
response options, partial correlations across the four LLMs ranged from
.43 to .49 for vulgarity, compared with -.12 to .14 for
inconsiderateness and -.27 to -.18 for lacking sense (Table S4). The
association with vulgarity also remained positive when inference
accounted for clustering by behavior (Table S5) and when each behavior
was removed in turn (Table S6).

\subsubsection{Does accuracy depend on the
society?}\label{does-accuracy-depend-on-the-society}

Because the question here is whether accuracy varies across societies
independently of general differences in rating level, we removed each
society's average level offset before measuring error. For each society,
we measured estimation error by the mean absolute error (MAE) across the
150 scenarios after removing that offset; Table 1 reports the unadjusted
error and the offset separately. After this adjustment, GPT-5's MAE
ranged from 0.60 to 1.34 scale points across societies; the other LLMs
showed similar variation. We examined whether this variation was
associated with societal development.

\paragraph{LLM estimates were somewhat more accurate for more developed
societies}\label{llm-estimates-were-somewhat-more-accurate-for-more-developed-societies}

With each society's offset removed, error fell with development in all
four LLMs (\emph{r} = -.52 to -.77; see Fig. 3). While this development
gradient is visually striking, it should be interpreted cautiously. More
developed societies tended to have less atypical norm profiles and
greater dispersion in their ratings across scenarios, and both profile
properties were associated with lower LLM error. After adjustment for
these properties, most, and sometimes all, of the development gradient
disappeared (Tables S7 and S8; Fig. S5).

\pandocbounded{\includegraphics[keepaspectratio]{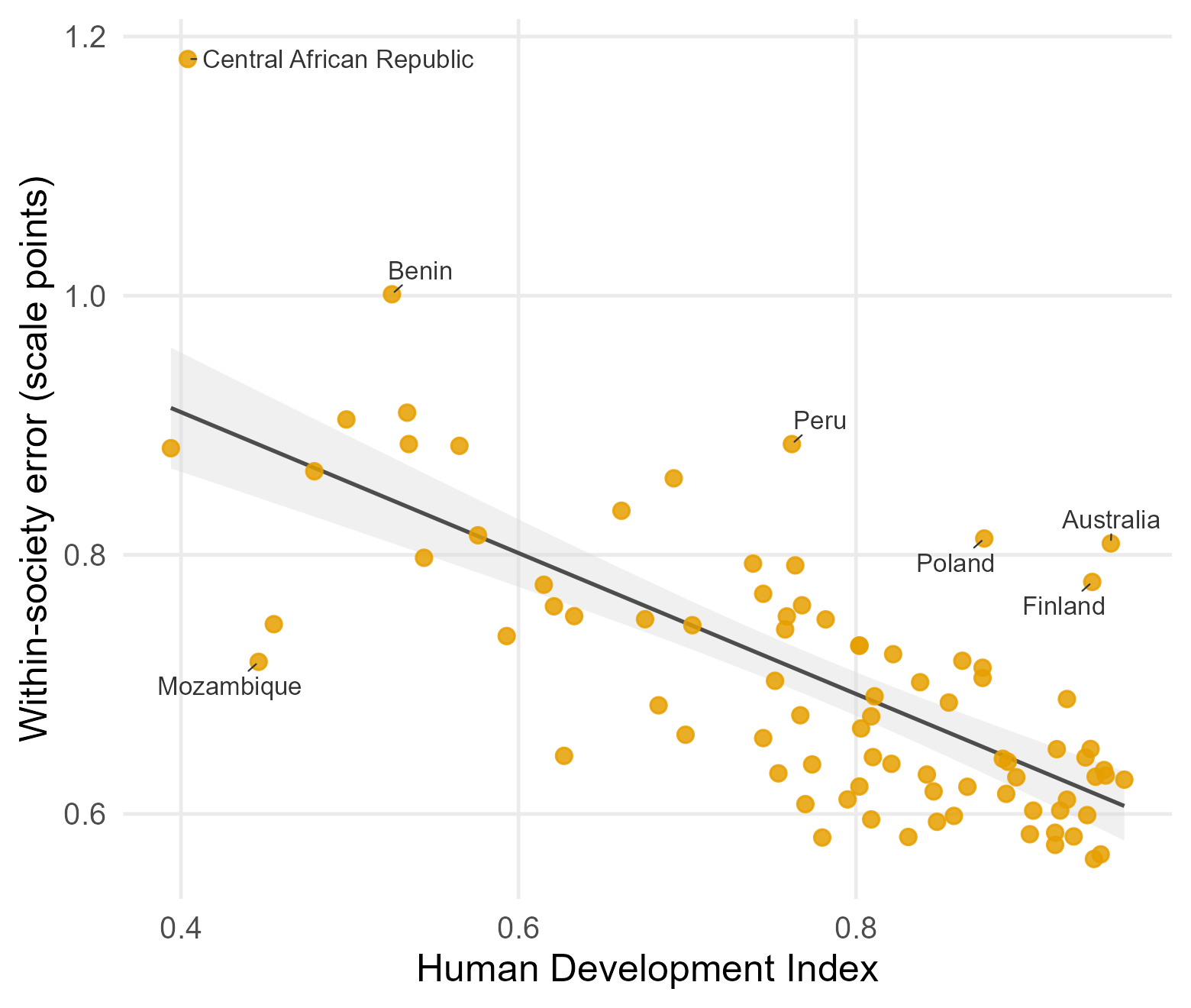}}

\textbf{Figure 3.} Within-society error against the Human Development
Index (HDI), averaged across the four LLMs. Each point is one of the 85
societies with available HDI value, and the plotted error is the mean of
the four LLMs' own within-society errors for that society, with each
society's level offset removed. The line is an ordinary least squares
fit with its 95\% confidence band, and we name the societies whose
errors differ from the fitted values by more than two residual standard
deviations. The error correlates with the index at \emph{r} = -.72 on
these averaged points, and the four LLMs' own correlations range from
\emph{r} = -.52 to -.77.

\subsubsection{How do alternative prompts affect the
findings?}\label{how-do-alternative-prompts-affect-the-findings}

We used several alternative prompts to examine whether the main findings
depended on the target population, prompt language, question wording, or
response scale. We separately omitted the society name to examine which
cultural profile the LLMs produced when no cultural target was
specified.

\emph{Prompting in local languages.} Because the main LLM collection
used English prompts whereas the GSEN survey had been administered in
local languages, we repeated the LLM task for 16 societies and 30
scenarios using the deposited GSEN translations. Local-language
prompting produced modest improvements: within-society error fell by
0.02 to 0.11 scale points, mean cross-cultural correlations rose by .04
to .12, and 22\% to 69\% of the compression shortfall under English
prompting was removed (Table S9). Nevertheless, every LLM continued to
estimate less between-society variation than the survey measured.

To determine whether the additional variation under local-language
prompting represented more accurate cultural differentiation, we crossed
four societies with four survey languages. The results were mixed:
compared with English prompting, the additional variation aligned more
closely with the surveyed differences in two LLMs but not in the other
two (Table S10). Thus, non-English prompts sometimes made the LLMs
differentiate societies more strongly, but not consistently more
accurately. Algeria and Saudi Arabia provided a particularly clean
comparison because the same Arabic survey instrument had been
administered in both societies. None of the four LLMs recovered the
surveyed difference between them under either English or Arabic
prompting (Table S9).

\emph{Changing the question and response scale.} For each LLM, we
collected two additional sets of estimates for 30 scenarios across all
90 societies. The scenarios comprised two randomly selected situations
for each of the 15 behaviors. Whereas the main prompt asked the LLM to
predict the average response that survey participants would give, one
alternative prompt asked directly how appropriate the behavior was in
the named society. The other retained the prediction task but replaced
the original -2.5 to 2.5 response scale with a 0-to-100 scale. Across
the main prompt and the two alternatives, compression remained below one
and the vulgarity gradient remained positive in all 12 LLM-by-prompt
combinations (Table S11). The compression finding was also robust to
weighting cells by respondent count, excluding sparsely sampled cells,
and correcting the spacing defect in the main prompt for a subset; under
these specifications, cross-cultural accuracy also remained far below
within-society accuracy (Tables S12 and S13).

\emph{Targeting university students.} We also examined whether weak
cross-cultural accuracy reflected a mismatch between the population
named in the prompt and the predominantly student population surveyed.
We evaluated 54 societies with sufficiently large student subsamples
(Methods). The student-only and full-sample society means were almost
identical, correlating at \emph{r} = .98 across societies for the
average scenario. Accordingly, estimates from the general prompt
correlated equally weakly with the full-sample and the student means, at
\emph{r} = .28 and \emph{r} = .28. The student prompt did change GPT-5's
estimates: across societies, the student-targeted and general-prompt
estimates correlated at only \emph{r} = .73 for the average scenario.
However, the student-targeted estimates correlated with the GSEN student
means at \emph{r} = .30, compared with \emph{r} = .28 for the
general-prompt estimates. Thus, matching the population named in the
prompt to the university-student population that predominated in the
survey samples produced little improvement.

\emph{Omitting the society name.} Finally, we asked each LLM to rate the
same 150 scenarios without naming a society, to address a separate
question: which cultural profile does an LLM produce when no cultural
target is specified? We compared each LLM's unconditioned profile with
its profiles for each of the 90 named societies. The profile for the
United States was closest to the unconditioned profile in three of the
four LLMs and second closest in the fourth, with distances ranging from
0.11 to 0.20 scale points. Canada, Australia, New Zealand and the United
Kingdom were also among the closest matches. The mean profile across all
named societies was further from the unconditioned profile, at 0.19 to
0.30 scale points (Table S14). The close match was not simply a
consequence of the United States being typical of the LLMs'
society-conditioned profiles: when those profiles were ranked by
distance from the mean of the other societies, the United States ranked
between 40 and 81 out of 90, with 1 denoting the closest profile. Nor
was the unconditioned profile distinctively close to the surveyed norms
of the United States. When the surveyed profiles were ranked by distance
from each LLM's unconditioned profile, the United States ranked between
4 and 24 out of 90, again with 1 denoting the closest profile. Thus, the
resemblance was specifically between the unconditioned profile and each
LLM's own representation of the United States.

\subsection{Discussion}\label{discussion}

The GSEN data allowed us to ask whether LLMs recover both the magnitude
and the pattern of societal differences across a common set of concrete
everyday scenarios. The LLMs represented cultural variation in everyday
norms poorly in both respects. First, they estimated differences in
norms between societies to be less than half their measured size.
Second, for many scenarios, they poorly identified the pattern of
cultural variation, that is, which societies judged the behavior less
acceptable and which judged it more acceptable.

The compression could arise because the LLMs possess little
society-specific information, or because the prompts fail to elicit or
calibrate information they possess. Our analyses do not distinguish
between these explanations. Prompting in local survey languages modestly
improved accuracy and increased cultural differentiation, but
substantial compression remained. This was not a general tendency to
compress numerical differences: within societies, the LLMs spread their
ratings across scenarios more widely than the survey did (Table S12).
The remaining shortfall may therefore reflect limited society-specific
knowledge, inadequate elicitation, poor numerical calibration, or some
combination.

Although the LLMs poorly identified the pattern of cultural variation
for many scenarios, they performed considerably better when the behavior
elicited concerns about vulgarity. One possible explanation is that
cultural differences involving vulgarity are discussed more explicitly
in text and are therefore more strongly represented in the data used to
develop LLMs (Adilazuarda et al., 2024; Atari et al., 2023). The
vulgarity marker has also been linked to purity, one of the binding
foundations in moral-foundations theory (Eriksson et al., 2025; Graham
et al., 2011). The result may therefore reflect better representation of
cultural differences involving vulgarity or associated purity concerns
than of differences involving many ordinary behaviors. This
interpretation remains provisional because the analysis covered only 15
behaviors.

Estimation errors were lower for more developed societies, consistent
with earlier reports of greater accuracy for Western or English-speaking
populations (Tao et al., 2024). This association should be interpreted
cautiously, however, because development was also associated with
properties of the surveyed norm profiles that predicted estimation
accuracy. The lower errors for developed societies may therefore reflect
those profile properties rather than better society-specific knowledge.
A second form of Western bias appeared when no society was named: the
unconditioned estimates most closely resembled those produced when the
prompts named the United States.

The study is limited by its human benchmark, which comprises unweighted
convenience-sample means rather than nationally representative
estimates. This limitation bears most directly on our finding that, for
many scenarios, the LLMs poorly identified which societies judged the
behavior less acceptable and which judged it more acceptable. Two
considerations address different aspects of this concern. First, the
GSEN sample means correspond closely to cross-society differences found
in representative surveys on independently measured values, including
belief in God and freedom of choice (Eriksson et al., 2025). This
correspondence supports using the GSEN means as a benchmark for
cross-cultural patterns, although nationally representative data on
everyday norms would provide a stronger test. Second, the most obvious
way in which the GSEN samples are unrepresentative is that they are
dominated by university students. However, the student-only and
full-sample society means were almost identical, and when an LLM was
explicitly asked about university students' everyday norms, it did not
identify the cultural pattern appreciably more accurately. Population
mismatch is therefore unlikely to provide a general explanation for the
poor representation of cultural variation.

We conclude that the four frontier LLMs tested here were not reliable
sources of information about how everyday norms differ between
societies. They substantially underestimated the size of cultural
differences and, for many scenarios, poorly identified their pattern.
Their greater accuracy for vulgarity-related behaviors shows that this
limitation is uneven rather than absolute.

\subsection{Materials and Methods}\label{materials-and-methods}

\subsubsection{Human benchmark}\label{human-benchmark}

\emph{Survey and sample.} The Global Study of Everyday Norms (GSEN;
Eriksson et al., 2025) surveyed 25,422 participants in 90 societies
between 14 July 2023 and 31 May 2024. Respondents answered in a local
language using one of 41 translated questionnaires. Across societies,
the mean student share was 70\% and the mean respondent age was 28. The
samples were not nationally representative. Three value measures
collected in the same survey correlated with corresponding measures from
nationally representative samples in the combined World Values Survey
and European Values Study dataset: \emph{r} = .73 across 73 societies
for freedom-of-choice values, .74 across 70 societies for
gender-egalitarian values, and .85 across 74 societies for belief in
God.

\emph{Appropriateness ratings.} The 150 scenarios combine each of 15
everyday behaviors with 10 situations. Participants rated scenario
appropriateness on a six-point scale from -2.5 to +2.5. We aggregated
the ratings into society-level means, based on a median of 54 ratings
per cell (range 1 to 217). The grid contains 13,468 of the 13,500
possible society-scenario cells. The kissing and flirting items were not
fielded in every society, and two additional scenarios received no
ratings in one society because of its small sample. For the 21 affected
scenarios, we calculated cross-cultural correlations across the
societies that rated them.

\emph{Vulgarity marker.} For up to three of the scenarios they rated,
randomly selected for each participant, participants also indicated what
someone who disapproves of the behavior would consider its main problem.
Three of the five response options name a moral concern: the behavior is
vulgar, inconsiderate, or lacks sense. The preregistered vulgarity
marker is the share of ``vulgar'' answers among these three concern
responses, pooled across respondents who answered the item for that
scenario.

\subsubsection{LLM estimation task}\label{llm-estimation-task}

We asked each LLM to estimate the mean rating that respondents in a
named society would give a scenario on the same -2.5 to +2.5 scale
(Supplementary Methods, Elicitation prompt). We queried the LLM versions
\texttt{gpt-5-2025-08-07}, \texttt{gpt-5.4-2026-03-05},
\texttt{claude-opus-4-6} and \texttt{gemini-3.1-pro-preview} five times
per society-scenario cell. For each cell, we averaged the valid
responses, defined as responses containing only a number; every cell had
at least one valid response. For each confirmatory LLM, we retried
failed calls up to three times to obtain five valid responses per cell.
Supplementary Methods (Valid responses) report the invalid responses.

We did not specify temperature or a top-p cutoff in any request, and our
requests did not explicitly enable extended reasoning. The four
endpoints do not offer equivalent settings, however, so the reasoning
configurations of the three confirmatory LLMs differed from that of
GPT-5. Supplementary Methods report the generation settings and
reasoning configuration used in each request. Differences in reasoning
settings limit comparisons of model capability.

Throughout the main data collection, the prompt omitted a space between
``of'' and the scenario. To assess whether this formatting defect
affected the results, we repeated data collection with the space
restored for 26 societies and 120 scenarios, twice per cell.

\subsubsection{Measures of accuracy}\label{measures-of-accuracy}

We compared LLM estimates with survey means along three dimensions: the
magnitude and the pattern of differences between societies, each
assessed for a scenario across the societies that rated it, and the
error within each society across its 150 scenarios. For each scenario,
compression is the ratio of the between-society SD of the LLM estimates
to the corresponding SD of the survey means, so that values below one
indicate underestimated differences between societies. Cross-cultural
accuracy is the Pearson \emph{r} between LLM estimates and survey means
across the societies that rated the scenario. This correlation is high
only when the LLM reproduces the pattern of differences between
societies for that scenario. Table 1 reports mean compression and mean
cross-cultural accuracy across scenarios.

Within each society, MAE is the mean absolute difference between LLM
estimates and survey means across the 150 scenarios, expressed in points
on the response scale. The level offset is the corresponding mean signed
difference, with negative values indicating that the LLM rates the
scenarios as less acceptable than respondents did. We calculated the
offset-adjusted MAE by subtracting the society's level offset from each
LLM-survey difference and then averaging the absolute values of the
adjusted differences. This adjustment removes the overall difference in
rating level between the LLM and respondents. Table 1 reports the
unadjusted within-society MAE and the level offset, each averaged across
societies. The society-level analyses use the offset-adjusted MAE.
Within-society agreement can also be measured using the Pearson
correlation (\emph{r}) between LLM estimates and survey means across the
150 scenarios. This correlation is unaffected by differences in their
overall level or spread and is reported in the supplement (Table S12 and
Fig. S2B).

Survey means and LLM averages both contain sampling error. We adjusted
for this error using a split-half approach. A single fixed split of
respondents within each society yielded two survey matrices. We formed
two LLM matrices by averaging the first two calls and the third and
fourth calls, respectively, omitting the fifth call from this
calculation to keep the halves equal in size. For each scenario, we
calculated separate split-half correlations for the survey and LLM
estimates, then used Spearman-Brown prophecy adjustments appropriate to
the full survey sample and the five-call LLM mean. To correct
cross-cultural accuracy for attenuation, we divided each scenario's
observed correlation by the square root of the product of these two
reliabilities. For the noise-corrected compression ratio, we estimated
the between-society SD of the survey means from the covariance between
the two survey matrices. We measured each society's sampling noise as
the mean absolute difference between the scenario means from its two
respondent halves, expressed in points on the response scale.

\subsubsection{Compression analysis}\label{compression-analysis}

The compression hypothesis was that the LLMs underestimate
between-society variation. The preregistered statistic is the unweighted
mean of compression across the 150 scenarios. We calculated its 95\%
percentile bootstrap interval from 2,000 resamples of scenarios. Under
the preregistered decision rule, the hypothesis is supported when the
entire interval lies below one. Because the ten scenarios of a behavior
are not independent, we instead resampled the 15 behaviors in a
sensitivity analysis and included all ten scenarios for each sampled
behavior. As a separate sensitivity analysis, we also calculated
compression using the noise-corrected survey SD.

\subsubsection{Vulgarity-gradient
analysis}\label{vulgarity-gradient-analysis}

The vulgarity hypothesis was that cross-cultural accuracy increases with
the vulgarity marker after adjustment for the between-society variation
available to predict. A larger between-society SD of survey means
indicates more cross-cultural variation for an LLM to estimate. Across
scenarios, we therefore regressed cross-cultural accuracy on the
vulgarity marker and the survey between-society SD, and we report the
corresponding partial correlation. The preregistered decision rule
requires a positive vulgarity coefficient with heteroskedasticity-robust
\emph{p} \textless{} .05, calculated using the HC3 estimator (MacKinnon
\& White, 1985).

Vulgarity differs mainly among the 15 behaviors rather than among their
10 situations. We therefore also report cluster-robust inference with
behaviors as clusters, using the bias-reduced linearization estimator
CR2 with Satterthwaite degrees of freedom (Pustejovsky \& Tipton, 2018).
For the confidence interval in Table 1, we resampled whole behaviors,
including all 10 situations for each. We also recalculated the partial
correlation after removing each behavior in turn for GPT-5.

The measure published by Eriksson et al.~(2025) divides the vulgar
answers by all answers across the five response options, whereas the
preregistered marker divides them by the three concern responses. We
repeated the preregistered test using the published proportion as a
sensitivity analysis. To examine whether the association was specific to
vulgarity, we calculated the same proportion of all answers for the
inconsiderate and lacks-sense responses and compared the partial
correlations of the three proportions with cross-cultural accuracy. We
used proportions of all answers in this comparison because three shares
that sum to one cannot all vary in the same direction. Each of the three
partial correlations controls for the survey between-society SD, as the
preregistered test does.

\subsubsection{Society-level analyses}\label{society-level-analyses}

We measured societal development using the HDI for the reference year
2021, published in the United Nations Development Programme's Human
Development Report 2021/22 (United Nations Development Programme, 2022,
Table 1). HDI values were available for 85 of the 90 societies. For each
LLM, we correlated development with the offset-adjusted MAE across these
societies. Because society-level means estimated from noisier samples
contribute more error to absolute measures of accuracy, we also
conducted a sensitivity analysis adjusting both development and
offset-adjusted MAE for each society's sampling noise (Table S7).

The association between development and accuracy could reflect
properties of the surveyed norm profiles rather than knowledge of the
societies themselves. We measured two such properties. A society's
atypicality is the mean absolute distance between its norm profile and
the average profile of the other societies, calculated scenario by
scenario with the society's own rating excluded from the average. As
with the offset-adjusted MAE, we first subtracted the society's mean
signed difference across scenarios from each scenario's difference, so
that atypicality reflects the pattern of the profile rather than its
overall level. A society's response dispersion is the standard deviation
of its survey means across the 150 scenarios. The dispersion measure
captures variation in a society's mean ratings across scenarios, rather
than disagreement among respondents within a scenario.

To test whether these two properties account for the association, we
regressed each society's offset-adjusted MAE on development, societal
atypicality and response dispersion simultaneously. We standardized the
outcome and all predictors and fitted a separate model for each LLM.

\subsubsection{Alternative prompt
conditions}\label{alternative-prompt-conditions}

The alternative prompt conditions varied the target population, prompt
language, question wording, and response scale. We ran the
university-student condition on GPT-5 alone and the other conditions on
all four LLMs. The no-society condition covered all 150 scenarios. Two
further conditions, each covering all 90 societies and 30 scenarios,
either asked directly how appropriate the behavior was in the named
society or replaced the response scale with a 0-to-100 scale. We
compared these two conditions and the language conditions with the main
collection restricted to the same cells and repetitions.

\emph{Local and crossed languages.} We conducted two language
comparisons with GSEN's deposited item translations. One prompted each
LLM about each society in the language of that society's survey
instrument, covering 16 societies and 30 scenarios. The other used a
crossed design that prompted each LLM about four societies (Brazil,
China, Germany, Turkey) in all four of their languages. We express the
reduction in compression as the share of the English-prompt shortfall
(one minus compression) that local-language prompting removed. To assess
whether added between-society variation matched the surveyed
differences, we regressed each LLM's deviations from the scenario mean
on the corresponding survey deviations.

\emph{University-student target.} We replaced ``people'' with
``university students'' in the main prompt, left its wording otherwise
unchanged, and collected GPT-5 estimates for all 150 scenarios in the 63
societies with at least one student respondent, five times per cell.
Before collecting these estimates, we prespecified the inclusion
criterion and restricted the analysis to the 54 societies whose student
respondents had been asked about each scenario at least 20 times on
average. We compared three pairings of estimates and society means, all
for the same cells: the student-targeted estimates with the means of the
GSEN student respondents alone, the general-prompt estimates from the
main GPT-5 collection with the full-sample means, and the same
general-prompt estimates with the student means. The third pairing
differs from the second only in the society means and from the first
only in the prompt. For each pairing, we calculated cross-cultural
accuracy for each scenario and averaged it across scenarios.

\emph{No society named.} We asked each LLM to rate every scenario five
times without specifying a society. For each LLM, we calculated mean
absolute distances from its unconditioned profile to each of its 90
society-conditioned profiles and to their mean. To assess typicality, we
ranked the society-conditioned profiles by their distance from the mean
profile of the other societies. For each LLM, we also ranked the 90
surveyed profiles by their distance from that LLM's unconditioned
profile. In both rankings, a rank of 1 indicated the smallest distance.

\subsubsection{Additional exploratory
analysis}\label{additional-exploratory-analysis}

Several exploratory analyses are reported only in the supplement. First,
we examined whether LLMs recovered changes in norms across two survey
periods. The comparison used the 26 societies and 120 scenarios shared
by GSEN, fielded in 2023--2024, and the earlier survey by Gelfand et
al.~(2011), fielded from 2000 to 2003. We compared the change implied by
each LLM's estimates for the present and for twenty years earlier with
the change measured between the two surveys (Table S15).

Two further analyses examined how well an LLM reproduced each society's
deviations from each scenario's cross-society mean. For each scenario,
we centered the LLM estimates and survey means separately by subtracting
their respective cross-society means. This centering removes differences
in average acceptability between scenarios, so that a deviation
indicates how much more or less acceptable than average a society finds
a scenario. For each society, we defined deviation accuracy as the
correlation between LLM and survey deviations across scenarios, and
deviation recovery as the slope from regressing LLM deviations on survey
deviations. We compared deviation recovery across four world regions
(West, Latin America, Asia and Africa; Table S16). We related deviation
accuracy to five society-level characteristics (Fig. S6) and examined
perceived tightness and the individualizing-binding axis in further
models (Tables S17 and S18).

We also compared each LLM with individual respondents. For each society,
we calculated the share of respondents whose own ratings differed more
from the mean ratings of the other respondents than the LLM estimates
differed from the society means, as measured by mean absolute
differences (Table S19). Finally, the supplement reports two of the
exploratory analyses listed in the confirmatory study's preregistration:
within-society accuracy by region (Table S20) and agreement between the
confirmatory LLMs in cross-cultural accuracy and estimated change across
scenarios (Table S21).

\subsubsection{Preregistration, data, and
code}\label{preregistration-data-and-code}

We used the exploratory GPT-5 analyses to formulate two hypotheses. H1
predicts that cross-cultural accuracy increases with the vulgarity
marker after adjustment for the between-society SD of survey means. H2
predicts that the mean ratio of LLM to survey between-society SD across
scenarios is below one. We tested both hypotheses in the other three
LLMs. All analyses other than these two confirmatory tests are
exploratory.

We preregistered the confirmatory study on OSF on April 13, 2026, before
querying GPT-5.4, Claude Opus 4.6 or Gemini 3.1 Pro
(https://osf.io/w3fxd). The preregistration specifies H1, H2 and their
decision rules, and also lists seven exploratory analyses without
specifying targets for them. For one of the exploratory analyses, the
preregistration specifies a within-society correlation as the measure of
agreement. In the main text, we instead report the within-society MAE
because it expresses the size of estimation errors in units of the
response scale. We report the preregistered correlation in the
supplement (Table S12 and Fig. S2B). The preregistration also states
that all models are run without reasoning or extended thinking. GPT-5.4
and Claude Opus 4.6 ran without reasoning, but Gemini 3.1 Pro returned
reasoning tokens on every call although its reasoning budget was set to
the smallest possible value (Supplementary Methods, Generation
settings). Both hypotheses were supported in all three LLMs, so neither
confirmatory result depends on this deviation.

The GPT-5 data collection was preregistered on OSF on September 11,
2025, before data collection began (https://osf.io/d7km9). That
preregistration specifies the cross-sectional benchmark, temporal task
and individual-level comparison (Table S19), but does not state any
hypotheses. We developed H1 and H2 from exploratory findings in that
dataset: a vulgarity gradient and compression of between-society
variation. Reporting the temporal statistics both with and without the
headphones scenarios was a post hoc decision for GPT-5 and was
pre-specified for the three confirmatory LLMs. After preregistering the
confirmatory study, we corrected the GPT-5 data because a parsing error
had left the stored rating column incomplete despite the responses
having been retained in the raw output. Supplementary Methods
(Correction of the exploratory GPT-5 collection) provide an audit of
this correction. The correction does not affect either confirmatory
decision rule, as both apply only to the other three LLMs.

Data, analysis code and full prompts are available via a view-only link
(\nolinkurl{https://osf.io/jzhqu/?view_only=4c217744f7564e8c8eef583e3ca6fd13}). The
archive will be made public upon acceptance.

\subsubsection{Ethical approval}\label{ethical-approval}

We analyzed previously published human-participant data from the Global
Study of Everyday Norms (GSEN) and collected new responses from large
language models. No new human-participant data were collected. The newly
collected data consisted entirely of AI-generated outputs and therefore
did not involve human subjects. No additional ethical approval was
required for the present study. Eriksson et al.~(2025) report the
ethical approvals for the original GSEN data collection.

\subsubsection{Use of AI}\label{use-of-ai}

We used AI tools to prepare the manuscript. We developed the analysis
code, including the code for tables and figures, through iterative work
in Claude Code over multiple sessions with Anthropic's Claude models:
successive versions of Claude Opus, most recently Claude Opus 5 and
Claude Opus 5.5, and Claude Sonnet 5. These models also drafted and
revised passages of text. To review drafts, identify possible
inconsistencies, and suggest editorial revisions, we used Claude Opus 5,
Claude Fable 5 and Claude Fable 5.1 from Anthropic, GPT-5.5, GPT-5.6 and
GPT-6 from OpenAI, and Gemini 3.1 Pro from Google. Throughout this
process, we directed the analyses and writing, critically evaluated the
outputs, and repeatedly revised the text; no AI-generated text was
incorporated without substantive human review. We take full
responsibility for the study design, analyses, interpretations,
conclusions, and final manuscript text.

\subsubsection{Funding}\label{funding}

This research was supported by the Knut and Alice Wallenberg Foundation
(grant no. 2022.0191).

\subsubsection{Competing interests}\label{competing-interests}

The authors declare no competing interests.

\subsection{References}\label{references}

Adilazuarda, M. F., Mukherjee, S., Lavania, P., Singh, S. S., Aji, A.
F., O'Neill, J., Modi, A., \& Choudhury, M. (2024). Towards measuring
and modeling ``culture'' in LLMs: A survey. \emph{Proceedings of EMNLP
2024}, 15763--15784. https://doi.org/10.18653/v1/2024.emnlp-main.882

Atari, M., Xue, M. J., Park, P. S., Blasi, D. E., \& Henrich, J. (2023).
Which humans? \emph{PsyArXiv}. https://doi.org/10.31234/osf.io/5b26t

Bulté, B., \& Rigouts Terryn, A. (2026). LLMs and cultural values: The
impact of prompt language and explicit cultural framing.
\emph{Computational Linguistics}, 52(2), 407--494.
https://doi.org/10.1162/coli.a.583

Eriksson, K., Karlsson, S., Vartanova, I., \& Strimling, P. (2026).
Large language models outperform humans at estimating society's everyday
norms. \emph{Communications AI \& Computing}, 1(1), 15.
https://doi.org/10.1038/s44488-026-00018-8

Eriksson, K., Strimling, P., Vartanova, I., \& Simpson, B. (2026). Same
flavours, different taste buds: A theory for predicting social norms for
specific behaviours across cultures. \emph{Journal of the Royal Society
Interface}, 23(237), 20251122. https://doi.org/10.1098/rsif.2025.1122

Eriksson, K., Strimling, P., Vartanova, I., Simpson, B., Persson, M., et
al.~(2025). Everyday norms have become more permissive over time and
vary across cultures. \emph{Communications Psychology}, 3(1), 66.
https://doi.org/10.1038/s44271-025-00324-4

Gelfand, M. J., Raver, J. L., Nishii, L., Leslie, L. M., Lun, J., et
al.~(2011). Differences between tight and loose cultures: A 33-nation
study. \emph{Science}, 332(6033), 1100--1104.
\nolinkurl{https://doi.org/10.1126/science.1197754}

Graham, J., Nosek, B. A., Haidt, J., Iyer, R., Koleva, S., \& Ditto, P.
H. (2011). Mapping the moral domain. \emph{Journal of Personality and
Social Psychology}, 101(2), 366--385. https://doi.org/10.1037/a0021847

MacKinnon, J. G., \& White, H. (1985). Some
heteroskedasticity-consistent covariance matrix estimators with improved
finite sample properties. \emph{Journal of Econometrics}, 29(3),
305--325. https://doi.org/10.1016/0304-4076(85)90158-7

Pustejovsky, J. E., \& Tipton, E. (2018). Small-sample methods for
cluster-robust variance estimation and hypothesis testing in fixed
effects models. \emph{Journal of Business \& Economic Statistics},
36(4), 672--683. https://doi.org/10.1080/07350015.2016.1247004

Ramezani, A., \& Xu, Y. (2023). Knowledge of cultural moral norms in
large language models. \emph{Proceedings of ACL 2023}, 428--446.
https://doi.org/10.18653/v1/2023.acl-long.26

Rao, A. S., Yerukola, A., Shah, V., Reinecke, K., \& Sap, M. (2025).
NormAd: A framework for measuring the cultural adaptability of large
language models. \emph{Proceedings of NAACL 2025}, 2373--2403.
https://doi.org/10.18653/v1/2025.naacl-long.120

Tao, Y., Viberg, O., Baker, R. S., \& Kizilcec, R. F. (2024). Cultural
bias and cultural alignment of large language models. \emph{PNAS Nexus},
3(9), pgae346. https://doi.org/10.1093/pnasnexus/pgae346

United Nations Development Programme. (2022). \emph{Human development
report 2021/22: Uncertain times, unsettled lives: Shaping our future in
a transforming world.} United Nations Development Programme.

Zhao, W., Mondal, D., Tandon, N., Dillion, D., Gray, K., \& Gu, Y.
(2024). WorldValuesBench: A large-scale benchmark dataset for
multi-cultural value awareness of language models. \emph{Proceedings of
LREC-COLING 2024}, 17696--17706.

\clearpage
\pdfbookmark[1]{Supplementary Information}{supplementary-information}
\begin{center}
{\LARGE Large language models underestimate and partly misrepresent
cultural variation in everyday norms\par}
\vskip 1em
{\Large Supplementary Information\par}
\end{center}
\vskip 1.5em

Kimmo Eriksson\textsuperscript{1,2}, Irina
Vartanova\textsuperscript{1,3}, Pontus Strimling\textsuperscript{1,4}

\textsuperscript{1}Institute for Futures Studies, Stockholm, Sweden\\
\textsuperscript{2}Department of Business and Mathematics, Mälardalen
University, Västerås, Sweden\\
\textsuperscript{3}Department of Women's and Children's Health, Uppsala
University, Uppsala, Sweden\\
\textsuperscript{4}Institute for Analytical Sociology, Linköping
University, Norrköping, Sweden

Correspondence: kimmo.eriksson@mdu.se

This Supplementary Information contains the Supplementary Methods,
Tables S1 to S21 and Figures S1 to S6 that the main text refers to.

\textbf{Contents}

\textbf{Supplementary Methods} --- Sample and grid; Elicitation prompt;
Generation settings; Valid responses; Correction of the exploratory
GPT-5 collection; Sampling noise and reliability; Deviation measures and
society-level predictors

\textbf{Supplementary Tables} --- Tables S1 to S21, in the order in
which the main text first refers to them

\textbf{Supplementary Figures} --- Figures S1 to S6, in the same order

\textbf{SI References} --- the works cited in this Supplementary
Information

\subsection{Supplementary Methods}\label{supplementary-methods}

\subsubsection{Sample and grid}\label{sample-and-grid}

The GSEN was administered online through Qualtrics except in Benin and
Mauritius, where interviewers administered the survey in person, across
125 sites recruiting through email, social media and survey
organizations, with compensation varying by site. Local collaborators
translated the survey from an English master version. The grid is
incomplete because the kissing items were not fielded in Kuwait or Saudi
Arabia, the flirting items were not fielded in Kuwait, and two further
cells (bargain in a restaurant and kiss in a public park) are absent for
Mongolia due to the small number of participants.

Analyses by region use the four world regions of the GSEN's own
classification: West with 38 societies, Latin America with 14 societies,
Asia with 22 societies, Africa with 16 societies. West is the widest of
the four and comprises all of Europe, together with Israel, Gibraltar,
Martinique, Canada, the United States, Australia and New Zealand.

\subsubsection{Elicitation prompt}\label{elicitation-prompt}

Each scenario-society rating was elicited with a single prompt, applied
verbatim across all four LLMs. The society name and the scenario phrase
were interpolated into the template below, which is reproduced here
exactly as the calls sent it.

\begin{quote}
From various sources in our everyday lives we have all developed a
subjective ``impression'' or ``feeling'' for the appropriateness of any
given behavior in a particular situation. Imagine a number of people
from {[}SOCIETY{]} rated the appropriateness of{[}SCENARIO{]} on the
scale from −2.5 through +2.5, where −2.5 = The behavior is extremely
inappropriate in this situation and +2.5 = The behavior is extremely
appropriate in this situation. Note that this scenario combines a
behavior and a situation (not a behavior and an object). Interpret it as
someone performing the behavior during the situation, in the way such a
phrase would typically be understood. For example, `writing on a bus'
means writing while riding the bus, not on its surface. Your task is to
estimate the average rating that these {[}SOCIETY{]} respondents would
give. Please use numbers with up to two decimals to provide as detailed
estimates as possible. Do not write any comments or justifications for
the estimates. Write only the estimate.
\end{quote}

The template joins ``the appropriateness of'' to the scenario phrase
without an intervening space, so every call reads ``the appropriateness
ofargue at a party''. The same template was used for the original GPT-5
collection and for all three confirmatory LLMs, so the wording is
constant across the main comparison.

Table S13 and the temporal comparison (Table S15) use a second
collection in which the space before the scenario is restored and
nothing else is changed. For the estimate twenty years earlier, its
prompt reads ``Imagine a number of people from {[}SOCIETY{]} twenty
years ago rated the appropriateness of {[}SCENARIO{]}'' and asks for the
average rating ``these {[}SOCIETY{]} respondents would have given twenty
years ago, say, in year 2003''.

\subsubsection{Generation settings}\label{generation-settings}

All four LLMs are closed-weight and were accessed through provider APIs.
The collection scripts specify the body of every request. No request
specified \texttt{temperature} or \texttt{top\_p}, and all requests used
an output cap of 2,048 tokens. Requests to both GPT models included the
model name, output cap, an empty system message and the prompt, but no
reasoning-effort field. Both therefore used the default reasoning effort
applied by their respective APIs. GPT-5.4 returned reasoning tokens on
0\% of calls, whereas GPT-5 returned them on 100\% of calls, with a
median of 320 reasoning tokens per call.

Requests to Claude Opus 4.6 included the model name, output cap and
prompt, but no extended-thinking field, leaving extended thinking
disabled by default. Its reasoning-token usage was not recorded. Gemini
3.1 Pro was the only LLM for which a reasoning field was explicitly set:
\texttt{thinkingBudget\ =\ 1}, the minimum accepted by the endpoint,
which rejects a budget of zero. Despite that setting, Gemini returned
reasoning tokens on 100\% of calls, with a median of 294 reasoning
tokens per call, the same rate as GPT-5. The confirmatory
preregistration states that ``All models are run without reasoning or
extended thinking''. Gemini's use of reasoning therefore constitutes a
deviation from the preregistration.

\subsubsection{Valid responses}\label{valid-responses}

All 195 refusals came from GPT-5, 0.24\% of its 82,935 calls in the
cross-sectional and temporal collections. A refusal declines in words
instead of giving a bare number, and ratings come from the bare-number
replies alone. Where a refusal contains a number inside the response
scale, the LLM has declined the society-specific question and given an
unconditioned estimate in its place, so that number answers a different
question. Another 84 GPT-5 calls returned no content (9 reached the
output limit, 1 returned an empty string without reaching it, and 74
were failed requests) and were not re-attempted, although the
preregistration of the exploratory study specifies up to three
re-attempts before a failed call is excluded. This deviation affects
0.10\% of GPT-5's calls, and every affected cell keeps 4 of its 5
repetitions. Every call of the three confirmatory LLMs returned a bare
number.

\subsubsection{Correction of the exploratory GPT-5
collection}\label{correction-of-the-exploratory-gpt-5-collection}

The exploratory GPT-5 collection was corrected after the replication was
preregistered. The preregistered hypotheses were developed from a
cross-cultural correlation of 0.209, a compression of 0.58 and a
vulgarity partial correlation of .464 for GPT-5. A parsing fault
produced those figures: it populated the stored rating column for only
15.6\% of the 67,340 cross-sectional calls, although the complete raw
reply was saved for every call. Where both exist they agree on 10,503 of
10,503 calls. The temporal grid had the same fault: the stored column
was populated for 20.0\% of its 15,595 calls, and the two values match
in all 3,126 of the 3,126 calls that have both. The corrected analysis
parses every rating from the saved raw reply. Neither preregistered
hypothesis names a numerical GPT-5 benchmark, so the correction does not
change either decision rule. H1 is reported using the three-answer
marker exactly as filed; the published all-five marker gives the same
conclusion as a sensitivity analysis.

\subsubsection{Sampling noise and
reliability}\label{sampling-noise-and-reliability}

The survey means that the LLMs are scored against are sample estimates,
and their precision enters three analyses that the main text reports and
this section details: the cross-cultural correlations corrected for
attenuation (Table S2), the noise-corrected compression (Table S1), and
the adjustment of the development gradient for each society's sampling
noise (Table S7). All three come from one split of the respondents.
Within each society, respondents are ranked and assigned alternately to
two halves, so the split is deterministic and a respondent stays in the
same half for every scenario they rated. Each half is aggregated into
its own society-by-scenario matrix of means. A cell whose few
respondents all fell into the same half has no pair, and the 36 such
cells, of 13,468, are left out of these calculations.

For each scenario, the two half-means are correlated across societies,
and the Spearman-Brown formula projects that correlation to the full
sample. The result is the reliability of the surveyed means, and its
square root is the ceiling: the correlation that a predictor equal to
the true society means would reach against the surveyed means. Averaged
over scenarios the ceiling is .90, so about a fifth of the
between-society variance in the surveyed means is respondent-sampling
noise. The LLM side is treated the same way. The mean of an LLM's first
two calls and the mean of its third and fourth are correlated across
societies, and the Spearman-Brown formula projects that correlation to
the five-call mean; the fifth call is left out so that the halves are
equal. Each scenario's observed correlation is then divided by the
square root of the product of the two reliabilities. A scenario drops
out of the corrected mean when the LLM's reliability cannot be estimated
as positive: the two halves correlate at zero or below, or the LLM gave
every society the same rating, which happens for a scenario the LLM
rates almost identically everywhere.

Sampling noise also inflates the survey's between-society SD, and with
it the denominator of compression. The covariance between the two
half-matrices, taken across the societies that rated a scenario, keeps
only the variation that both halves share; its square root is the
between-society SD without respondent-sampling noise, which replaces the
survey SD in the noise-corrected compression of Table S1.

A society's sampling noise is the mean absolute gap between the scenario
means of its two halves, in points on the response scale. Table S7
partials it out of the correlations of error with development and with
atypicality, because societies with smaller samples have noisier means
and sampling noise is related to both.

\subsubsection{Deviation measures and society-level
predictors}\label{deviation-measures-and-society-level-predictors}

Two measures score how well an LLM reproduces a society's deviations
from each scenario's cross-society mean. Deviation accuracy is the
correlation, within a society across scenarios, between the LLM's and
the survey's deviations from each scenario's cross-society mean.
Deviation recovery is the slope from regressing, within a society across
scenarios, the LLM's deviations on the survey's deviations.

A society's position on the individualizing-binding axis has three
measures. The first is the GSEN's nine-item forced-choice scale, on
which a respondent chooses between an individualizing and a binding
violation on each item. A society's vulgarity weight is the random
intercept from a logistic mixed model with society and scenario random
intercepts. The foundation score is the society mean of the nine binding
items in the separate 17-item endorsement battery. Perceived tightness
is the society mean of the six-item tightness scale of Gelfand et
al.~(2011). Society strictness is the society's mean negative-sanction
rating across the 150 scenarios, the negative of its mean
appropriateness rating.

Predictors of accuracy are tested with OLS regression, and region is a
four-level factor with West as reference.

\subsection{Supplementary Tables}\label{supplementary-tables}

\needspace{8\baselineskip}
\textbf{Table S1.} Compression under alternative computations.

\begin{longtable}[]{@{}
  >{\raggedright\arraybackslash}p{(\linewidth - 6\tabcolsep) * \real{0.3068}}
  >{\raggedright\arraybackslash}p{(\linewidth - 6\tabcolsep) * \real{0.2444}}
  >{\raggedright\arraybackslash}p{(\linewidth - 6\tabcolsep) * \real{0.2145}}
  >{\raggedright\arraybackslash}p{(\linewidth - 6\tabcolsep) * \real{0.2343}}@{}}
\toprule\noalign{}
\begin{minipage}[b]{\linewidth}\raggedright
LLM
\end{minipage} & \begin{minipage}[b]{\linewidth}\raggedright
Compression
\end{minipage} & \begin{minipage}[b]{\linewidth}\raggedright
95\% interval, behaviors resampled
\end{minipage} & \begin{minipage}[b]{\linewidth}\raggedright
Noise-corrected compression
\end{minipage} \\
\midrule\noalign{}
\endhead
\bottomrule\noalign{}
\endlastfoot
GPT-5 & 0.49 & {[}0.38, 0.62{]} & 0.54 \\
GPT-5.4 & 0.43 & {[}0.34, 0.53{]} & 0.47 \\
Claude Opus 4.6 & 0.47 & {[}0.38, 0.57{]} & 0.52 \\
Gemini 3.1 Pro & 0.47 & {[}0.36, 0.60{]} & 0.51 \\
\end{longtable}

\emph{Note.} Compression is as defined in Table 1 of the main text. The
interval is a 95\% percentile interval from 2,000 bootstrap resamples of
the 15 behaviors, each resample keeping all the scenarios of every
behavior it draws. Noise-corrected compression replaces the survey SD
with the between-society SD without respondent-sampling noise
(Supplementary Methods, Sampling noise and reliability).

\needspace{8\baselineskip}
\textbf{Table S2.} Cross-cultural accuracy corrected for measurement
precision.

\begin{longtable}[]{@{}llll@{}}
\toprule\noalign{}
LLM & Cross-cultural r & LLM reliability & Corrected r \\
\midrule\noalign{}
\endhead
\bottomrule\noalign{}
\endlastfoot
GPT-5 & 0.27 & 0.77 & 0.32 \\
GPT-5.4 & 0.24 & 0.59 & 0.33 \\
Claude Opus 4.6 & 0.23 & 0.94 & 0.26 \\
Gemini 3.1 Pro & 0.33 & 0.76 & 0.40 \\
\end{longtable}

\emph{Note.} Each scenario's cross-cultural correlation is divided by
the square root of the product of the survey reliability and the LLM
reliability, and the corrected values are averaged (Supplementary
Methods, Sampling noise and reliability). LLM reliability is the
reliability of the five-call mean. The survey reliability implies a
ceiling of .90, the same for every LLM. A scenario is dropped from the
corrected mean when the LLM's reliability cannot be estimated as
positive; this removed three scenarios for GPT-5, one each for GPT-5.4
and Claude Opus 4.6, and none for Gemini 3.1 Pro.

\needspace{8\baselineskip}
\textbf{Table S3.} Behavior-level cross-cultural accuracy and vulgarity
marker for GPT-5.

\begin{longtable}[]{@{}
  >{\raggedright\arraybackslash}p{(\linewidth - 4\tabcolsep) * \real{0.4000}}
  >{\raggedright\arraybackslash}p{(\linewidth - 4\tabcolsep) * \real{0.2267}}
  >{\raggedright\arraybackslash}p{(\linewidth - 4\tabcolsep) * \real{0.3733}}@{}}
\toprule\noalign{}
\begin{minipage}[b]{\linewidth}\raggedright
Behavior
\end{minipage} & \begin{minipage}[b]{\linewidth}\raggedright
Cross-cultural r
\end{minipage} & \begin{minipage}[b]{\linewidth}\raggedright
Registered vulgarity marker
\end{minipage} \\
\midrule\noalign{}
\endhead
\bottomrule\noalign{}
\endlastfoot
kiss (on the mouth) & 0.62 & 0.55 \\
flirt & 0.58 & 0.46 \\
curse or swear & 0.42 & 0.46 \\
eat & 0.33 & 0.16 \\
listen to music on headphones & 0.31 & 0.10 \\
laugh out loud & 0.28 & 0.16 \\
sing & 0.28 & 0.10 \\
shout in anger & 0.23 & 0.20 \\
rest & 0.20 & 0.11 \\
cry & 0.17 & 0.07 \\
argue & 0.16 & 0.15 \\
use a mobile phone & 0.12 & 0.08 \\
bargain & 0.12 & 0.13 \\
read the newspaper & 0.11 & 0.08 \\
talk & 0.07 & 0.13 \\
\end{longtable}

\emph{Note.} Each entry averages the 10 scenarios of a behavior.
Cross-cultural \emph{r} is as defined in Table 1, with each scenario's
correlation taken across the 87 to 90 societies that rated it
(Supplementary Methods, Sample and grid). The registered vulgarity
marker is the preregistered marker defined in Materials and Methods; it
comes from the survey and is therefore the same for every LLM.

\needspace{8\baselineskip}
\textbf{Table S4.} Partial correlations of cross-cultural accuracy with
the vulgarity marker and with alternative concerns.

\begin{longtable}[]{@{}
  >{\raggedright\arraybackslash}p{(\linewidth - 8\tabcolsep) * \real{0.1739}}
  >{\raggedright\arraybackslash}p{(\linewidth - 8\tabcolsep) * \real{0.1739}}
  >{\raggedright\arraybackslash}p{(\linewidth - 8\tabcolsep) * \real{0.1522}}
  >{\raggedright\arraybackslash}p{(\linewidth - 8\tabcolsep) * \real{0.2609}}
  >{\raggedright\arraybackslash}p{(\linewidth - 8\tabcolsep) * \real{0.2391}}@{}}
\toprule\noalign{}
\begin{minipage}[b]{\linewidth}\raggedright
LLM
\end{minipage} & \begin{minipage}[b]{\linewidth}\raggedright
Registered + SD
\end{minipage} & \begin{minipage}[b]{\linewidth}\raggedright
All five + SD
\end{minipage} & \begin{minipage}[b]{\linewidth}\raggedright
Inconsiderate, given SD
\end{minipage} & \begin{minipage}[b]{\linewidth}\raggedright
Lacks sense, given SD
\end{minipage} \\
\midrule\noalign{}
\endhead
\bottomrule\noalign{}
\endlastfoot
GPT-5 & 0.52 & 0.49 & 0.12 & -0.25 \\
GPT-5.4 & 0.51 & 0.43 & -0.12 & -0.27 \\
Claude Opus 4.6 & 0.54 & 0.48 & -0.01 & -0.20 \\
Gemini 3.1 Pro & 0.48 & 0.46 & 0.14 & -0.18 \\
\end{longtable}

\emph{Note.} Entries are partial correlations between cross-cultural
accuracy and one concern proportion over the 150 scenarios, each
controlling for the survey between-society SD. For Claude Opus 4.6, one
scenario received the same rating in every society, so its
cross-cultural correlation is undefined and is scored as zero. The
concern item offered five response options: the behavior is vulgar, is
inconsiderate, lacks sense, other, or the respondent cannot imagine that
anyone would disapprove. The first column uses the preregistered marker,
which divides the vulgar answers by the answers to the first three
options, and repeats the vulgarity partial correlation of Table 1. The
other three columns divide the vulgar, the inconsiderate and the
lacks-sense answers by the answers to all five options; the vulgar
proportion is the one published by Eriksson et al.~(2025).

\needspace{8\baselineskip}
\textbf{Table S5.} The vulgarity gradient under
heteroskedasticity-robust and behavior-clustered inference.

\begin{longtable}[]{@{}
  >{\raggedright\arraybackslash}p{(\linewidth - 16\tabcolsep) * \real{0.1882}}
  >{\raggedright\arraybackslash}p{(\linewidth - 16\tabcolsep) * \real{0.2235}}
  >{\raggedright\arraybackslash}p{(\linewidth - 16\tabcolsep) * \real{0.0588}}
  >{\raggedright\arraybackslash}p{(\linewidth - 16\tabcolsep) * \real{0.0824}}
  >{\raggedright\arraybackslash}p{(\linewidth - 16\tabcolsep) * \real{0.0824}}
  >{\raggedright\arraybackslash}p{(\linewidth - 16\tabcolsep) * \real{0.0824}}
  >{\raggedright\arraybackslash}p{(\linewidth - 16\tabcolsep) * \real{0.0471}}
  >{\raggedright\arraybackslash}p{(\linewidth - 16\tabcolsep) * \real{0.1529}}
  >{\raggedright\arraybackslash}p{(\linewidth - 16\tabcolsep) * \real{0.0824}}@{}}
\toprule\noalign{}
\begin{minipage}[b]{\linewidth}\raggedright
LLM
\end{minipage} & \begin{minipage}[b]{\linewidth}\raggedright
Predictor
\end{minipage} & \begin{minipage}[b]{\linewidth}\raggedright
b
\end{minipage} & \begin{minipage}[b]{\linewidth}\raggedright
HC3 SE
\end{minipage} & \begin{minipage}[b]{\linewidth}\raggedright
HC3 p
\end{minipage} & \begin{minipage}[b]{\linewidth}\raggedright
CR2 SE
\end{minipage} & \begin{minipage}[b]{\linewidth}\raggedright
df
\end{minipage} & \begin{minipage}[b]{\linewidth}\raggedright
CR2 95\% CI
\end{minipage} & \begin{minipage}[b]{\linewidth}\raggedright
CR2 p
\end{minipage} \\
\midrule\noalign{}
\endhead
\bottomrule\noalign{}
\endlastfoot
GPT-5 & Between-society SD & 0.64 & 0.11 & \textless{} .001 & 0.17 & 9.8
& {[}0.25, 1.02{]} & .004 \\
GPT-5 & Vulgarity & 0.78 & 0.11 & \textless{} .001 & 0.08 & 4.0 &
{[}0.56, 1.00{]} & \textless{} .001 \\
GPT-5.4 & Between-society SD & 0.61 & 0.09 & \textless{} .001 & 0.12 &
9.8 & {[}0.35, 0.87{]} & \textless{} .001 \\
GPT-5.4 & Vulgarity & 0.67 & 0.09 & \textless{} .001 & 0.08 & 4.0 &
{[}0.44, 0.90{]} & .001 \\
Claude Opus 4.6 & Between-society SD & 0.58 & 0.07 & \textless{} .001 &
0.09 & 9.8 & {[}0.39, 0.78{]} & \textless{} .001 \\
Claude Opus 4.6 & Vulgarity & 0.71 & 0.09 & \textless{} .001 & 0.09 &
4.0 & {[}0.45, 0.97{]} & .002 \\
Gemini 3.1 Pro & Between-society SD & 0.60 & 0.09 & \textless{} .001 &
0.13 & 9.8 & {[}0.31, 0.89{]} & .001 \\
Gemini 3.1 Pro & Vulgarity & 0.66 & 0.09 & \textless{} .001 & 0.11 & 4.0
& {[}0.35, 0.97{]} & .004 \\
\end{longtable}

\emph{Note.} Entries come from one ordinary least squares regression per
LLM, across the 150 scenarios, of cross-cultural accuracy on the
preregistered vulgarity marker and the survey between-society SD; Table
1 gives the partial correlations from the same regressions. The
coefficients are unstandardized: the outcome is a correlation, the
marker a proportion and the SD in scale points. The two estimators share
b and differ only in the standard error and \emph{p} value. The H1
decision rule uses the HC3 \emph{p} on the vulgarity rows of the three
confirmatory LLMs; GPT-5 is shown for reference. The CR2 standard errors
cluster on the 15 behaviors, and for vulgarity the Satterthwaite degrees
of freedom are about four, which widens the CR2 interval.

\needspace{8\baselineskip}
\textbf{Table S6.} Vulgarity gradient for GPT-5 with each behavior
removed in turn.

\begin{longtable}[]{@{}lll@{}}
\toprule\noalign{}
Behavior dropped & Partial r & Vulgar b \\
\midrule\noalign{}
\endhead
\bottomrule\noalign{}
\endlastfoot
kiss (on the mouth) & 0.46 & 0.79 \\
curse or swear & 0.49 & 0.80 \\
flirt & 0.49 & 0.79 \\
laugh out loud & 0.51 & 0.76 \\
use a mobile phone & 0.52 & 0.75 \\
shout in anger & 0.52 & 0.76 \\
cry & 0.52 & 0.78 \\
rest & 0.52 & 0.77 \\
argue & 0.52 & 0.78 \\
bargain & 0.53 & 0.76 \\
talk & 0.53 & 0.74 \\
read the newspaper & 0.53 & 0.77 \\
eat & 0.53 & 0.79 \\
sing & 0.55 & 0.82 \\
listen to music on headphones & 0.55 & 0.82 \\
\end{longtable}

\emph{Note.} Each row drops the 10 scenarios of one behavior and refits
the GPT-5 regression of Table S5 on the remaining 140 scenarios; partial
r is the vulgarity partial correlation and vulgar b the unstandardized
vulgarity coefficient.

\needspace{8\baselineskip}
\textbf{Table S7.} Correlations of offset-adjusted within-society MAE
with development and societal atypicality.

\begin{longtable}[]{@{}
  >{\raggedright\arraybackslash}p{(\linewidth - 8\tabcolsep) * \real{0.2515}}
  >{\raggedright\arraybackslash}p{(\linewidth - 8\tabcolsep) * \real{0.2030}}
  >{\raggedright\arraybackslash}p{(\linewidth - 8\tabcolsep) * \real{0.2030}}
  >{\raggedright\arraybackslash}p{(\linewidth - 8\tabcolsep) * \real{0.1713}}
  >{\raggedright\arraybackslash}p{(\linewidth - 8\tabcolsep) * \real{0.1712}}@{}}
\toprule\noalign{}
\begin{minipage}[b]{\linewidth}\raggedright
LLM
\end{minipage} & \begin{minipage}[b]{\linewidth}\raggedright
Development r
\end{minipage} & \begin{minipage}[b]{\linewidth}\raggedright
Development r, adjusted for sampling noise
\end{minipage} & \begin{minipage}[b]{\linewidth}\raggedright
Atypicality r
\end{minipage} & \begin{minipage}[b]{\linewidth}\raggedright
Atypicality r, adjusted for sampling noise
\end{minipage} \\
\midrule\noalign{}
\endhead
\bottomrule\noalign{}
\endlastfoot
GPT-5 & -0.75 & -0.75 & 0.41 & 0.39 \\
GPT-5.4 & -0.52 & -0.49 & 0.56 & 0.53 \\
Claude Opus 4.6 & -0.66 & -0.63 & 0.64 & 0.60 \\
Gemini 3.1 Pro & -0.77 & -0.76 & 0.45 & 0.42 \\
\end{longtable}

\emph{Note.} Entries are Pearson correlations across the 85 societies
with an available Human Development Index; atypicality is defined in
Materials and Methods, Society-level analyses. The adjusted columns
partial each society's sampling noise, as defined in Materials and
Methods, out of both the predictor and the error; sampling noise
correlates with development at \emph{r} = -.25 and with atypicality at
\emph{r} = .41.

\needspace{8\baselineskip}
\textbf{Table S8.} Development, societal atypicality and response
dispersion as predictors of offset-adjusted within-society MAE.

\begin{longtable}[]{@{}
  >{\raggedright\arraybackslash}p{(\linewidth - 8\tabcolsep) * \real{0.1644}}
  >{\raggedright\arraybackslash}p{(\linewidth - 8\tabcolsep) * \real{0.2055}}
  >{\raggedright\arraybackslash}p{(\linewidth - 8\tabcolsep) * \real{0.2055}}
  >{\raggedright\arraybackslash}p{(\linewidth - 8\tabcolsep) * \real{0.2192}}
  >{\raggedright\arraybackslash}p{(\linewidth - 8\tabcolsep) * \real{0.2055}}@{}}
\toprule\noalign{}
\begin{minipage}[b]{\linewidth}\raggedright
Predictor
\end{minipage} & \begin{minipage}[b]{\linewidth}\raggedright
GPT-5
\end{minipage} & \begin{minipage}[b]{\linewidth}\raggedright
GPT-5.4
\end{minipage} & \begin{minipage}[b]{\linewidth}\raggedright
Claude Opus 4.6
\end{minipage} & \begin{minipage}[b]{\linewidth}\raggedright
Gemini 3.1 Pro
\end{minipage} \\
\midrule\noalign{}
\endhead
\bottomrule\noalign{}
\endlastfoot
Development & -0.14 (.030) & 0.15 (.132) & -0.05 (.519) & -0.17
(.004) \\
Atypicality & 0.22 (\textless{} .001) & 0.49 (\textless{} .001) & 0.51
(\textless{} .001) & 0.25 (\textless{} .001) \\
Dispersion & -0.75 (\textless{} .001) & -0.69 (\textless{} .001) & -0.59
(\textless{} .001) & -0.72 (\textless{} .001) \\
\end{longtable}

\emph{Note.} Cells give the standardized coefficient, with its \emph{p}
value in parentheses, from one ordinary least squares model per LLM
across the 85 societies with an available Human Development Index.
Dispersion and development correlate at \emph{r} = .71 across these
societies, so the model cannot cleanly separate their contributions;
atypicality is less confounded with development (\emph{r} = -.37).

\needspace{8\baselineskip}
\textbf{Table S9.} Local-language prompts compared with English prompts
in the 16 societies whose survey translations the GSEN archive provides.

\begin{longtable}[]{@{}
  >{\raggedright\arraybackslash}p{(\linewidth - 14\tabcolsep) * \real{0.1655}}
  >{\raggedright\arraybackslash}p{(\linewidth - 14\tabcolsep) * \real{0.1596}}
  >{\raggedright\arraybackslash}p{(\linewidth - 14\tabcolsep) * \real{0.0924}}
  >{\raggedright\arraybackslash}p{(\linewidth - 14\tabcolsep) * \real{0.1580}}
  >{\raggedright\arraybackslash}p{(\linewidth - 14\tabcolsep) * \real{0.0826}}
  >{\raggedright\arraybackslash}p{(\linewidth - 14\tabcolsep) * \real{0.0645}}
  >{\raggedright\arraybackslash}p{(\linewidth - 14\tabcolsep) * \real{0.1722}}
  >{\raggedright\arraybackslash}p{(\linewidth - 14\tabcolsep) * \real{0.1052}}@{}}
\toprule\noalign{}
\begin{minipage}[b]{\linewidth}\raggedright
LLM
\end{minipage} & \begin{minipage}[b]{\linewidth}\raggedright
Prompt language
\end{minipage} & \begin{minipage}[b]{\linewidth}\raggedright
Cross-cultural r
\end{minipage} & \begin{minipage}[b]{\linewidth}\raggedright
Compression
\end{minipage} & \begin{minipage}[b]{\linewidth}\raggedright
Aimed slope
\end{minipage} & \begin{minipage}[b]{\linewidth}\raggedright
Error
\end{minipage} & \begin{minipage}[b]{\linewidth}\raggedright
English − local
\end{minipage} & \begin{minipage}[b]{\linewidth}\raggedright
Shortfall removed
\end{minipage} \\
\midrule\noalign{}
\endhead
\bottomrule\noalign{}
\endlastfoot
GPT-5 & English & 0.235 & 0.46 & 0.40 & 0.90 & 0.09 {[}0.00, 0.20{]}
& \\
GPT-5 & Local language & 0.330 & 0.58 & 0.47 & 0.81 & & 23\% \\
GPT-5.4 & English & 0.145 & 0.41 & 0.28 & 0.90 & 0.11 {[}-0.02, 0.26{]}
& \\
GPT-5.4 & Local language & 0.268 & 0.72 & 0.38 & 0.79 & & 53\% \\
Claude Opus 4.6 & English & 0.242 & 0.41 & 0.33 & 0.66 & 0.02 {[}-0.07,
0.10{]} & \\
Claude Opus 4.6 & Local language & 0.285 & 0.82 & 0.45 & 0.64 & &
69\% \\
Gemini 3.1 Pro & English & 0.304 & 0.47 & 0.45 & 0.88 & 0.04 {[}-0.03,
0.13{]} & \\
Gemini 3.1 Pro & Local language & 0.427 & 0.58 & 0.57 & 0.84 & & 22\% \\
\end{longtable}

\emph{Note.} Cross-cultural r and compression are as in Table 1,
computed on the condition's cells. Aimed slope is the slope from
regressing LLM deviations on survey deviations, each centered on its own
scenario mean across the 16 societies; a slope of 1 would recover the
surveyed between-society differences in full. Error is the unadjusted
within-society MAE of Table 1. English − local, on each LLM's English
row, is the paired reduction in error under local-language prompting,
with a 95\% interval from the behavior bootstrap of Table S1. Shortfall
removed, on the local-language row, is the share of the English prompt's
compression shortfall, one minus compression, that local-language
prompting removes.

The 16 societies are those for which the GSEN archive provides item
translations in a language used in that society's survey: Algeria and
Saudi Arabia (Arabic), Brazil (Portuguese), China (Simplified Chinese),
Iran (Persian), Israel (Hebrew, male version), Mexico (Spanish), Rwanda
(Kinyarwanda), and France, Germany, Greece, Japan, Poland, South Korea,
Turkey and Vietnam in their national languages. The analysis used 30
scenarios, two from each of the 15 behaviors, and the framing sentences
were machine-translated and checked by machine back-translation. Both
conditions have two calls per cell; the English rows are the main
collection restricted to the same cells and its first two repetitions.
Regressing each LLM's estimated difference between Algeria and Saudi
Arabia on the surveyed difference, across the scenarios both rated,
gives slopes of -.02 to .10 in English and -.19 to .06 in Arabic.

\needspace{8\baselineskip}
\textbf{Table S10.} Crossed design: four societies prompted in each of
their four languages.

\begin{longtable}[]{@{}
  >{\raggedright\arraybackslash}p{(\linewidth - 12\tabcolsep) * \real{0.1203}}
  >{\raggedright\arraybackslash}p{(\linewidth - 12\tabcolsep) * \real{0.1579}}
  >{\raggedright\arraybackslash}p{(\linewidth - 12\tabcolsep) * \real{0.1729}}
  >{\raggedright\arraybackslash}p{(\linewidth - 12\tabcolsep) * \real{0.0902}}
  >{\raggedright\arraybackslash}p{(\linewidth - 12\tabcolsep) * \real{0.1504}}
  >{\raggedright\arraybackslash}p{(\linewidth - 12\tabcolsep) * \real{0.1504}}
  >{\raggedright\arraybackslash}p{(\linewidth - 12\tabcolsep) * \real{0.1579}}@{}}
\toprule\noalign{}
\begin{minipage}[b]{\linewidth}\raggedright
LLM
\end{minipage} & \begin{minipage}[b]{\linewidth}\raggedright
Prompt language
\end{minipage} & \begin{minipage}[b]{\linewidth}\raggedright
Repeatable compression
\end{minipage} & \begin{minipage}[b]{\linewidth}\raggedright
Aimed slope
\end{minipage} & \begin{minipage}[b]{\linewidth}\raggedright
Own − EN
\end{minipage} & \begin{minipage}[b]{\linewidth}\raggedright
Fixed − EN
\end{minipage} & \begin{minipage}[b]{\linewidth}\raggedright
Aimed, own − EN (SE)
\end{minipage} \\
\midrule\noalign{}
\endhead
\bottomrule\noalign{}
\endlastfoot
GPT-5 & English & 0.47 & 0.35 & & & \\
GPT-5 & Brazilian Portuguese & 0.45 & 0.32 & & & \\
GPT-5 & German & 0.50 & 0.29 & & & \\
GPT-5 & Simplified Chinese & 0.38 & 0.26 & & & \\
GPT-5 & Turkish & 0.51 & 0.15 & & & \\
GPT-5 & Own language & 0.46 & 0.28 & -0.01 {[}-0.20, 0.20{]} & -0.01
{[}-0.09, 0.05{]} & -0.07 (0.08) \\
GPT-5.4 & English & 0.36 & 0.21 & & & \\
GPT-5.4 & Brazilian Portuguese & 0.36 & 0.37 & & & \\
GPT-5.4 & German & 0.48 & 0.39 & & & \\
GPT-5.4 & Simplified Chinese & 0.26 & 0.31 & & & \\
GPT-5.4 & Turkish & 0.32 & 0.23 & & & \\
GPT-5.4 & Own language & 0.75 & 0.36 & 0.39 {[}0.07, 0.72{]} & 0.00
{[}-0.29, 0.20{]} & 0.16 (0.16) \\
Claude Opus 4.6 & English & 0.41 & 0.40 & & & \\
Claude Opus 4.6 & Brazilian Portuguese & 0.58 & 0.40 & & & \\
Claude Opus 4.6 & German & 0.71 & 0.26 & & & \\
Claude Opus 4.6 & Simplified Chinese & 0.55 & 0.25 & & & \\
Claude Opus 4.6 & Turkish & 0.74 & 0.24 & & & \\
Claude Opus 4.6 & Own language & 0.80 & 0.41 & 0.39 {[}0.13, 0.68{]} &
0.23 {[}0.08, 0.40{]} & 0.00 (0.11) \\
Gemini 3.1 Pro & English & 0.43 & 0.48 & & & \\
Gemini 3.1 Pro & Brazilian Portuguese & 0.54 & 0.53 & & & \\
Gemini 3.1 Pro & German & 0.53 & 0.50 & & & \\
Gemini 3.1 Pro & Simplified Chinese & 0.48 & 0.50 & & & \\
Gemini 3.1 Pro & Turkish & 0.50 & 0.36 & & & \\
Gemini 3.1 Pro & Own language & 0.65 & 0.53 & 0.22 {[}0.06, 0.44{]} &
0.09 {[}-0.01, 0.19{]} & 0.05 (0.05) \\
\end{longtable}

\emph{Note.} Brazil, China, Germany and Turkey are each prompted in each
of the four societies' survey languages. A fixed-language row prompts
all four societies in that language; the Own language row prompts each
society in its own survey language, and the English row comes from the
main collection, both as in Table S9. All six configurations use the 28
scenarios with complete data for every LLM and configuration; the other
two scenarios of Table S9 are excluded because one LLM returned ratings
for fewer than four societies in one language. The aimed slope is as in
Table S9. Repeatable compression estimates the LLM's between-society SD
from the covariance between the configuration's two repetitions, so it
keeps only the variation shared by both; the survey SD in the
denominator still includes sampling noise. Own − EN is the own-language
minus the English repeatable compression, and Fixed − EN is the mean of
the four fixed-language values minus the English value. Both are shown
once per LLM, on its Own language row, with 95\% intervals from the
behavior bootstrap of Table S1 over the behaviors among the 28
scenarios. Within one language the four societies share whatever that
language does to the response scale, so Fixed − EN is the part of the
increase that does not come from giving each society a different
language. Aimed, own − EN is the change in the aimed slope from the
English to the own-language configuration, estimated as an interaction
term in one regression over both configurations, with the standard error
clustered by scenario in parentheses.

\needspace{8\baselineskip}
\textbf{Table S11.} Main statistics under the original prompt, the
direct question and the 0-to-100 response scale.

\begin{longtable}[]{@{}
  >{\raggedright\arraybackslash}p{(\linewidth - 8\tabcolsep) * \real{0.2366}}
  >{\raggedright\arraybackslash}p{(\linewidth - 8\tabcolsep) * \real{0.2171}}
  >{\raggedright\arraybackslash}p{(\linewidth - 8\tabcolsep) * \real{0.1105}}
  >{\raggedright\arraybackslash}p{(\linewidth - 8\tabcolsep) * \real{0.1885}}
  >{\raggedright\arraybackslash}p{(\linewidth - 8\tabcolsep) * \real{0.2473}}@{}}
\toprule\noalign{}
\begin{minipage}[b]{\linewidth}\raggedright
LLM
\end{minipage} & \begin{minipage}[b]{\linewidth}\raggedright
Prompt
\end{minipage} & \begin{minipage}[b]{\linewidth}\raggedright
Cross-cultural r
\end{minipage} & \begin{minipage}[b]{\linewidth}\raggedright
Compression
\end{minipage} & \begin{minipage}[b]{\linewidth}\raggedright
Vulgarity partial r {[}95\% CI{]}
\end{minipage} \\
\midrule\noalign{}
\endhead
\bottomrule\noalign{}
\endlastfoot
GPT-5 & Original & 0.244 & 0.40 & 0.46 {[}-0.05, 0.69{]} \\
GPT-5 & Direct question & 0.269 & 0.48 & 0.46 {[}-0.06, 0.69{]} \\
GPT-5 & 0-100 scale & 0.258 & 0.41 & 0.50 {[}-0.01, 0.73{]} \\
GPT-5.4 & Original & 0.208 & 0.41 & 0.43 {[}-0.06, 0.66{]} \\
GPT-5.4 & Direct question & 0.308 & 0.46 & 0.36 {[}-0.11, 0.58{]} \\
GPT-5.4 & 0-100 scale & 0.219 & 0.42 & 0.41 {[}-0.24, 0.72{]} \\
Claude Opus 4.6 & Original & 0.254 & 0.46 & 0.50 {[}0.18, 0.70{]} \\
Claude Opus 4.6 & Direct question & 0.289 & 0.55 & 0.25 {[}-0.11,
0.50{]} \\
Claude Opus 4.6 & 0-100 scale & 0.276 & 0.47 & 0.40 {[}0.09, 0.62{]} \\
Gemini 3.1 Pro & Original & 0.261 & 0.39 & 0.50 {[}0.07, 0.69{]} \\
Gemini 3.1 Pro & Direct question & 0.256 & 0.52 & 0.39 {[}-0.06,
0.63{]} \\
Gemini 3.1 Pro & 0-100 scale & 0.329 & 0.42 & 0.42 {[}-0.04, 0.64{]} \\
\end{longtable}

\emph{Note.} ``Direct question'' asks for the rating without the framing
sentence about imagined respondents; ``0-100 scale'' moves the response
scale and is mapped back to the original units for comparison. Every LLM
and every condition rated the same 30 scenarios, two of the ten
situations of each behavior, drawn at random in the collection script
under a single fixed seed; the original-prompt rows are the main
collection cut to these scenarios and to its first two repetitions.
Cross-cultural r and compression are as in Table 1. The interval on the
vulgarity gradient is a 95\% interval from the behavior bootstrap of
Table S1, with the same resampled behaviors applied to all three
prompts.

\needspace{8\baselineskip}
\textbf{Table S12.} Main statistics with respondent-count weighting and
with sparsely sampled cells excluded.

{\footnotesize
\begin{longtable}[]{@{}
  >{\raggedright\arraybackslash}p{(\linewidth - 14\tabcolsep) * \real{0.1298}}
  >{\raggedright\arraybackslash}p{(\linewidth - 14\tabcolsep) * \real{0.2917}}
  >{\raggedright\arraybackslash}p{(\linewidth - 14\tabcolsep) * \real{0.0766}}
  >{\raggedright\arraybackslash}p{(\linewidth - 14\tabcolsep) * \real{0.0762}}
  >{\raggedright\arraybackslash}p{(\linewidth - 14\tabcolsep) * \real{0.1301}}
  >{\raggedright\arraybackslash}p{(\linewidth - 14\tabcolsep) * \real{0.1369}}
  >{\raggedright\arraybackslash}p{(\linewidth - 14\tabcolsep) * \real{0.0689}}
  >{\raggedright\arraybackslash}p{(\linewidth - 14\tabcolsep) * \real{0.0897}}@{}}
\toprule\noalign{}
\begin{minipage}[b]{\linewidth}\raggedright
LLM
\end{minipage} & \begin{minipage}[b]{\linewidth}\raggedright
Variant
\end{minipage} & \begin{minipage}[b]{\linewidth}\raggedright
Within-society r
\end{minipage} & \begin{minipage}[b]{\linewidth}\raggedright
Cross-cultural r
\end{minipage} & \begin{minipage}[b]{\linewidth}\raggedright
Compression
\end{minipage} & \begin{minipage}[b]{\linewidth}\raggedright
Amplification
\end{minipage} & \begin{minipage}[b]{\linewidth}\raggedright
Cells
\end{minipage} & \begin{minipage}[b]{\linewidth}\raggedright
Societies
\end{minipage} \\
\midrule\noalign{}
\endhead
\bottomrule\noalign{}
\endlastfoot
GPT-5 & a. unweighted, all cells & 0.86 & 0.27 & 0.49 & 1.58 & 13,468 &
90 \\
GPT-5 & b. weighted by cell n & 0.86 & 0.26 & 0.54 & 1.58 & 13,468 &
90 \\
GPT-5 & c.~drop cells with n \textless{} 20 & 0.86 & 0.28 & 0.53 & 1.58
& 12,193 & 83 \\
GPT-5 & d.~drop societies with median cell n \textless{} 30 & 0.86 &
0.27 & 0.53 & 1.60 & 11,380 & 76 \\
GPT-5.4 & a. unweighted, all cells & 0.83 & 0.24 & 0.43 & 1.42 & 13,468
& 90 \\
GPT-5.4 & b. weighted by cell n & 0.83 & 0.24 & 0.46 & 1.42 & 13,468 &
90 \\
GPT-5.4 & c.~drop cells with n \textless{} 20 & 0.84 & 0.25 & 0.45 &
1.42 & 12,193 & 83 \\
GPT-5.4 & d.~drop societies with median cell n \textless{} 30 & 0.84 &
0.25 & 0.45 & 1.44 & 11,380 & 76 \\
Claude Opus 4.6 & a. unweighted, all cells & 0.83 & 0.23 & 0.47 & 1.33 &
13,468 & 90 \\
Claude Opus 4.6 & b. weighted by cell n & 0.83 & 0.24 & 0.50 & 1.33 &
13,468 & 90 \\
Claude Opus 4.6 & c.~drop cells with n \textless{} 20 & 0.84 & 0.25 &
0.50 & 1.33 & 12,193 & 83 \\
Claude Opus 4.6 & d.~drop societies with median cell n \textless{} 30 &
0.84 & 0.24 & 0.50 & 1.34 & 11,380 & 76 \\
Gemini 3.1 Pro & a. unweighted, all cells & 0.87 & 0.33 & 0.47 & 1.51 &
13,468 & 90 \\
Gemini 3.1 Pro & b. weighted by cell n & 0.87 & 0.32 & 0.52 & 1.51 &
13,468 & 90 \\
Gemini 3.1 Pro & c.~drop cells with n \textless{} 20 & 0.88 & 0.33 &
0.50 & 1.51 & 12,193 & 83 \\
Gemini 3.1 Pro & d.~drop societies with median cell n \textless{} 30 &
0.88 & 0.33 & 0.51 & 1.53 & 11,380 & 76 \\
\end{longtable}
}

\emph{Note.} Variant a is the main analysis. Within-society r is the
correlation between LLM estimates and survey means across a society's
scenarios, averaged over societies (Materials and Methods, Measures of
accuracy); the within-society r of variant a is the preregistered
measure of agreement that Materials and Methods, Preregistration, data,
and code, refers to. Amplification is the ratio of the LLM to the survey
SD across a society's scenarios, averaged over societies. Cross-cultural
r and compression are as in Table 1. Variant b weights each society by
the number of respondents behind its cell mean when a statistic runs
across societies, and each scenario in the same way when a statistic
runs across scenarios. Variant c drops the cells with fewer than 20
respondents; variant d drops the societies with a median of fewer than
30 respondents per cell and keeps every cell of the others.

\needspace{8\baselineskip}
\textbf{Table S13.} Main statistics under the fielded prompt and the
prompt with corrected spacing.

\begin{longtable}[]{@{}
  >{\raggedright\arraybackslash}p{(\linewidth - 10\tabcolsep) * \real{0.1924}}
  >{\raggedright\arraybackslash}p{(\linewidth - 10\tabcolsep) * \real{0.3130}}
  >{\raggedright\arraybackslash}p{(\linewidth - 10\tabcolsep) * \real{0.0903}}
  >{\raggedright\arraybackslash}p{(\linewidth - 10\tabcolsep) * \real{0.0898}}
  >{\raggedright\arraybackslash}p{(\linewidth - 10\tabcolsep) * \real{0.1533}}
  >{\raggedright\arraybackslash}p{(\linewidth - 10\tabcolsep) * \real{0.1612}}@{}}
\toprule\noalign{}
\begin{minipage}[b]{\linewidth}\raggedright
LLM
\end{minipage} & \begin{minipage}[b]{\linewidth}\raggedright
Prompt
\end{minipage} & \begin{minipage}[b]{\linewidth}\raggedright
Within-society r
\end{minipage} & \begin{minipage}[b]{\linewidth}\raggedright
Cross-cultural r
\end{minipage} & \begin{minipage}[b]{\linewidth}\raggedright
Compression
\end{minipage} & \begin{minipage}[b]{\linewidth}\raggedright
Amplification
\end{minipage} \\
\midrule\noalign{}
\endhead
\bottomrule\noalign{}
\endlastfoot
GPT-5 & Fielded, 5 repetitions & .89 & .30 & .68 & 1.49 \\
GPT-5 & Fielded, repetitions 1 and 2 & .89 & .28 & .72 & 1.49 \\
GPT-5 & Fielded, repetitions 3 and 4 & .89 & .29 & .71 & 1.49 \\
GPT-5 & Repaired, 2 repetitions & .89 & .29 & .73 & 1.49 \\
GPT-5.4 & Fielded, 5 repetitions & .86 & .26 & .53 & 1.38 \\
GPT-5.4 & Fielded, repetitions 1 and 2 & .86 & .23 & .65 & 1.40 \\
GPT-5.4 & Fielded, repetitions 3 and 4 & .86 & .21 & .63 & 1.39 \\
GPT-5.4 & Repaired, 2 repetitions & .87 & .22 & .66 & 1.42 \\
Claude Opus 4.6 & Fielded, 5 repetitions & .88 & .26 & .58 & 1.26 \\
Claude Opus 4.6 & Fielded, repetitions 1 and 2 & .87 & .26 & .61 &
1.26 \\
Claude Opus 4.6 & Fielded, repetitions 3 and 4 & .88 & .25 & .60 &
1.26 \\
Claude Opus 4.6 & Repaired, 2 repetitions & .88 & .26 & .66 & 1.23 \\
Gemini 3.1 Pro & Fielded, 5 repetitions & .91 & .39 & .63 & 1.43 \\
Gemini 3.1 Pro & Fielded, repetitions 1 and 2 & .91 & .37 & .68 &
1.43 \\
Gemini 3.1 Pro & Fielded, repetitions 3 and 4 & .91 & .36 & .67 &
1.43 \\
Gemini 3.1 Pro & Repaired, 2 repetitions & .91 & .38 & .67 & 1.43 \\
\end{longtable}

\emph{Note.} All rows use the 26 societies and 120 scenarios of the
collection with corrected spacing, the grid shared with the earlier
survey of Gelfand et al.~(2011); because that grid is a subset of the
main grid, the statistics differ from their Table 1 values. The repaired
condition restores the space the fielded prompt omits before the
scenario (Supplementary Methods, Elicitation prompt), a fault every
other condition of the study inherits, and changes nothing else; it has
two calls per cell. The first row of each LLM is the main collection on
these cells. The two rows below it are disjoint pairs of its
repetitions, with the same measurement noise as the repaired condition,
so they show how much two calls vary under identical wording.
Within-society r, cross-cultural r, compression and amplification are
defined in the note to Table S12.

\needspace{8\baselineskip}
\textbf{Table S14.} The no-society condition: each LLM's unconditioned
profile against the mean of its society-conditioned profiles, and the
named society it lies closest to.

\begin{longtable}[]{@{}lll@{}}
\toprule\noalign{}
LLM & MAE & Closest society distance \\
\midrule\noalign{}
\endhead
\bottomrule\noalign{}
\endlastfoot
GPT-5 & 0.19 & 0.11 (United States) \\
GPT-5.4 & 0.30 & 0.20 (United States) \\
Claude Opus 4.6 & 0.22 & 0.16 (Australia) \\
Gemini 3.1 Pro & 0.23 & 0.11 (United States) \\
\end{longtable}

\emph{Note.} MAE is the mean absolute distance across the 150 scenarios,
in points on the response scale, between an LLM's unconditioned profile
and the mean of its profiles for the 90 named societies. The last column
gives the named society whose profile lies closest to the unconditioned
profile, with that distance; in Claude Opus 4.6, where the closest is
Australia, the United States ranks second at 0.17 points. For scale, an
LLM's profile for a single named society lies on average 0.19 to 0.23
points from the same mean profile, and its society-conditioned answers
have between-society SDs averaging 0.25 to 0.29 points in the main
collection.

\needspace{8\baselineskip}
\textbf{Table S15.} Accuracy of the estimated change in norms between
the two surveys.

\begin{longtable}[]{@{}
  >{\raggedright\arraybackslash}p{(\linewidth - 12\tabcolsep) * \real{0.2127}}
  >{\raggedleft\arraybackslash}p{(\linewidth - 12\tabcolsep) * \real{0.1127}}
  >{\raggedright\arraybackslash}p{(\linewidth - 12\tabcolsep) * \real{0.1483}}
  >{\raggedright\arraybackslash}p{(\linewidth - 12\tabcolsep) * \real{0.1434}}
  >{\raggedright\arraybackslash}p{(\linewidth - 12\tabcolsep) * \real{0.1344}}
  >{\raggedright\arraybackslash}p{(\linewidth - 12\tabcolsep) * \real{0.1222}}
  >{\raggedright\arraybackslash}p{(\linewidth - 12\tabcolsep) * \real{0.1263}}@{}}
\toprule\noalign{}
\begin{minipage}[b]{\linewidth}\raggedright
LLM
\end{minipage} & \begin{minipage}[b]{\linewidth}\raggedleft
Scenarios
\end{minipage} & \begin{minipage}[b]{\linewidth}\raggedright
Aggregate r {[}95\% CI{]}
\end{minipage} & \begin{minipage}[b]{\linewidth}\raggedright
Magnitude slope {[}95\% CI{]}
\end{minipage} & \begin{minipage}[b]{\linewidth}\raggedright
Directional accuracy {[}95\% CI{]}
\end{minipage} & \begin{minipage}[b]{\linewidth}\raggedright
Loosening baseline
\end{minipage} & \begin{minipage}[b]{\linewidth}\raggedright
Within-society mean r {[}95\% CI{]}
\end{minipage} \\
\midrule\noalign{}
\endhead
\bottomrule\noalign{}
\endlastfoot
GPT-5 (all scenarios) & 120 & 0.243 {[}0.067, 0.405{]} & 0.12 {[}0.03,
0.21{]} & 52\% {[}43, 62{]} & 56\% & 0.132 {[}0.082, 0.181{]} \\
GPT-5 (headphones removed) & 110 & -0.082 {[}-0.265, 0.107{]} & -0.03
{[}-0.09, 0.04{]} & 50\% {[}40, 59{]} & 51\% & 0.008 {[}-0.045,
0.062{]} \\
GPT-5.4 (all scenarios) & 120 & -0.002 {[}-0.181, 0.177{]} & -0.00
{[}-0.13, 0.13{]} & 46\% {[}37, 56{]} & 56\% & -0.004 {[}-0.060,
0.053{]} \\
GPT-5.4 (headphones removed) & 110 & -0.089 {[}-0.272, 0.100{]} & -0.06
{[}-0.20, 0.07{]} & 45\% {[}35, 55{]} & 51\% & -0.030 {[}-0.083,
0.022{]} \\
Claude Opus 4.6 (all scenarios) & 120 & 0.227 {[}0.050, 0.390{]} & 0.12
{[}0.03, 0.21{]} & 50\% {[}41, 60{]} & 56\% & 0.131 {[}0.090,
0.173{]} \\
Claude Opus 4.6 (headphones removed) & 110 & 0.005 {[}-0.183, 0.192{]} &
0.00 {[}-0.08, 0.09{]} & 46\% {[}36, 56{]} & 51\% & 0.036 {[}-0.007,
0.079{]} \\
Gemini 3.1 Pro (all scenarios) & 120 & 0.462 {[}0.308, 0.592{]} & 0.23
{[}0.15, 0.31{]} & 65\% {[}56, 74{]} & 56\% & 0.213 {[}0.166,
0.261{]} \\
Gemini 3.1 Pro (headphones removed) & 110 & 0.345 {[}0.169, 0.500{]} &
0.08 {[}0.04, 0.13{]} & 62\% {[}52, 71{]} & 51\% & 0.087 {[}0.037,
0.138{]} \\
\end{longtable}

\emph{Note.} Measured change is the GSEN mean minus the mean of the
earlier survey (Gelfand et al., 2011), over the 26 societies and 120
scenarios the two surveys share, and estimated change is an LLM's
present-day estimate minus its estimate for twenty years earlier. The
surveys sampled different respondents twenty years apart, so the task
evaluates retrospective estimates of the difference between two survey
waves, not change followed in one population over time, and a difference
between the waves includes changes in the instruments and sample
composition as well as changes in norms. Both estimates come from the
collection with corrected prompt spacing (Supplementary Methods,
Elicitation prompt), with two calls per cell at each time point, and all
four LLMs answered the same cells.

Aggregate r is the correlation across scenarios between measured and
estimated change, after each has been averaged over societies.
Within-society mean r is the correlation between measured and estimated
change across scenarios within each society, averaged over societies.
Magnitude slope is the slope from regressing estimated change on
measured change using the same scenario means as aggregate r.
Directional accuracy is the proportion of scenarios for which estimated
and measured change have the same sign. This calculation includes only
scenarios with an absolute measured change greater than 0.01 scale
points, because a scenario that did not change has no direction to
predict. The loosening baseline is the proportion of those same
scenarios that became more permissive, corresponding to the accuracy of
predicting loosening in every scenario.

Every statistic is reported with and without the 10 headphones
scenarios, given previously documented changes in appropriateness as
personal audio devices became ubiquitous. All intervals are 95\%
intervals, obtained from the correlation test for aggregate r,
regression for the magnitude slope, the binomial test for directional
accuracy, and a t interval across societies for within-society mean r.

\needspace{8\baselineskip}
\textbf{Table S16.} Deviation recovery by region.

\begin{longtable}[]{@{}
  >{\raggedright\arraybackslash}p{(\linewidth - 8\tabcolsep) * \real{0.1800}}
  >{\raggedright\arraybackslash}p{(\linewidth - 8\tabcolsep) * \real{0.1967}}
  >{\raggedright\arraybackslash}p{(\linewidth - 8\tabcolsep) * \real{0.2300}}
  >{\raggedright\arraybackslash}p{(\linewidth - 8\tabcolsep) * \real{0.1967}}
  >{\raggedright\arraybackslash}p{(\linewidth - 8\tabcolsep) * \real{0.1966}}@{}}
\toprule\noalign{}
\begin{minipage}[b]{\linewidth}\raggedright
LLM
\end{minipage} & \begin{minipage}[b]{\linewidth}\raggedright
West (n = 38)
\end{minipage} & \begin{minipage}[b]{\linewidth}\raggedright
Latin Am. (n = 14)
\end{minipage} & \begin{minipage}[b]{\linewidth}\raggedright
Asia (n = 22)
\end{minipage} & \begin{minipage}[b]{\linewidth}\raggedright
Africa (n = 16)
\end{minipage} \\
\midrule\noalign{}
\endhead
\bottomrule\noalign{}
\endlastfoot
GPT-5 & 0.25 {[}0.22, 0.29{]} & 0.28 {[}0.19, 0.36{]} & 0.38
{[}0.26, 0.51{]} & 0.25 {[}0.15, 0.35{]} \\
GPT-5.4 & 0.18 {[}0.14, 0.22{]} & 0.17 {[}0.10, 0.23{]} & 0.27
{[}0.16, 0.38{]} & 0.18 {[}0.09, 0.26{]} \\
Claude Opus 4.6 & 0.18 {[}0.14, 0.21{]} & 0.19 {[}0.11, 0.26{]} & 0.26 {[}0.16, 0.37{]} & 0.24 {[}0.16, 0.32{]} \\
Gemini 3.1 Pro & 0.29 {[}0.25, 0.33{]} & 0.24 {[}0.14, 0.33{]}
& 0.44 {[}0.31, 0.56{]} & 0.33 {[}0.22, 0.44{]} \\
\end{longtable}

\emph{Note.} Cells give the mean recovery slope and its 95\% \emph{t}
interval, and each column head gives the number of societies in the
region. Deviation recovery is defined in
Supplementary Methods, Deviation measures and society-level predictors,
and the four regions in Sample and grid.

\needspace{8\baselineskip}
\textbf{Table S17.} Perceived tightness as a predictor of deviation
accuracy under six specifications.

\begin{longtable}[]{@{}
  >{\raggedright\arraybackslash}p{(\linewidth - 8\tabcolsep) * \real{0.1702}}
  >{\raggedright\arraybackslash}p{(\linewidth - 8\tabcolsep) * \real{0.5426}}
  >{\raggedleft\arraybackslash}p{(\linewidth - 8\tabcolsep) * \real{0.1064}}
  >{\raggedright\arraybackslash}p{(\linewidth - 8\tabcolsep) * \real{0.1277}}
  >{\raggedright\arraybackslash}p{(\linewidth - 8\tabcolsep) * \real{0.0532}}@{}}
\toprule\noalign{}
\begin{minipage}[b]{\linewidth}\raggedright
LLM
\end{minipage} & \begin{minipage}[b]{\linewidth}\raggedright
Specification
\end{minipage} & \begin{minipage}[b]{\linewidth}\raggedleft
Societies
\end{minipage} & \begin{minipage}[b]{\linewidth}\raggedright
Coefficient
\end{minipage} & \begin{minipage}[b]{\linewidth}\raggedright
p
\end{minipage} \\
\midrule\noalign{}
\endhead
\bottomrule\noalign{}
\endlastfoot
GPT-5 & Tightness alone & 89 & 0.24 & .027 \\
GPT-5 & Plus strictness and ground-truth quality & 89 & 0.22 & .044 \\
GPT-5 & The same, on the societies with sample composition & 71 & 0.20 &
.097 \\
GPT-5 & Plus student and female shares of the sample & 71 & 0.08 &
.537 \\
GPT-5 & Plus region & 89 & 0.20 & .161 \\
GPT-5 & Gelfand et al.~(2011) tightness, alone & 26 & 0.26 & .193 \\
GPT-5.4 & Tightness alone & 89 & 0.13 & .211 \\
GPT-5.4 & Plus strictness and ground-truth quality & 89 & 0.12 & .288 \\
GPT-5.4 & The same, on the societies with sample composition & 71 & 0.07
& .537 \\
GPT-5.4 & Plus student and female shares of the sample & 71 & -0.08 &
.522 \\
GPT-5.4 & Plus region & 89 & 0.09 & .541 \\
GPT-5.4 & Gelfand et al.~(2011) tightness, alone & 26 & 0.08 & .692 \\
Claude Opus 4.6 & Tightness alone & 89 & 0.17 & .096 \\
Claude Opus 4.6 & Plus strictness and ground-truth quality & 89 & 0.15 &
.143 \\
Claude Opus 4.6 & The same, on the societies with sample composition &
71 & 0.17 & .151 \\
Claude Opus 4.6 & Plus student and female shares of the sample & 71 &
0.05 & .662 \\
Claude Opus 4.6 & Plus region & 89 & 0.12 & .382 \\
Claude Opus 4.6 & Gelfand et al.~(2011) tightness, alone & 26 & 0.10 &
.639 \\
Gemini 3.1 Pro & Tightness alone & 89 & 0.29 & .006 \\
Gemini 3.1 Pro & Plus strictness and ground-truth quality & 89 & 0.27 &
.010 \\
Gemini 3.1 Pro & The same, on the societies with sample composition & 71
& 0.29 & .016 \\
Gemini 3.1 Pro & Plus student and female shares of the sample & 71 &
0.17 & .165 \\
Gemini 3.1 Pro & Plus region & 89 & 0.18 & .198 \\
Gemini 3.1 Pro & Gelfand et al.~(2011) tightness, alone & 26 & 0.26 &
.205 \\
\end{longtable}

\emph{Note.} Entries are the coefficient of perceived tightness and its
ordinary p value from one ordinary least squares regression per LLM and
specification, fitted across societies with deviation accuracy as the
outcome. Every variable is standardized within LLM, so coefficients are
comparable down a column and across specifications. Deviation accuracy,
tightness, strictness and region are defined in Supplementary Methods,
Deviation measures and society-level predictors. Ground-truth quality is
how closely an average respondent agrees with their own society's means:
for each respondent who rated at least 30 scenarios, the correlation
between their ratings and the leave-one-out mean of the other
respondents of the same society and scenario, averaged over the
society's respondents. The fourth and fifth specifications add their
named terms to the second. The 89-society specifications omit Kuwait,
which has no score on the GSEN tightness scale. The third refits the
second on the 71 societies for which the student share is published, so
that the fourth differs from it by the composition controls alone. The
last scores the same six-item scale on the respondents Gelfand et
al.~(2011) surveyed two decades earlier, in 26 of these societies, and
is standardized on those societies alone, so its coefficient is the
correlation on that subset; those scores correlate with the GSEN
tightness scale at \emph{r} = .71.

\needspace{8\baselineskip}
\textbf{Table S18.} Society-level predictors of deviation recovery under
three reference distributions.

\begin{longtable}[]{@{}
  >{\raggedright\arraybackslash}p{(\linewidth - 10\tabcolsep) * \real{0.2115}}
  >{\raggedright\arraybackslash}p{(\linewidth - 10\tabcolsep) * \real{0.3572}}
  >{\raggedright\arraybackslash}p{(\linewidth - 10\tabcolsep) * \real{0.0839}}
  >{\raggedright\arraybackslash}p{(\linewidth - 10\tabcolsep) * \real{0.0766}}
  >{\raggedright\arraybackslash}p{(\linewidth - 10\tabcolsep) * \real{0.1162}}
  >{\raggedright\arraybackslash}p{(\linewidth - 10\tabcolsep) * \real{0.1546}}@{}}
\toprule\noalign{}
\begin{minipage}[b]{\linewidth}\raggedright
LLM
\end{minipage} & \begin{minipage}[b]{\linewidth}\raggedright
Predictor
\end{minipage} & \begin{minipage}[b]{\linewidth}\raggedright
Slope
\end{minipage} & \begin{minipage}[b]{\linewidth}\raggedright
Naive p
\end{minipage} & \begin{minipage}[b]{\linewidth}\raggedright
Region-clustered p
\end{minipage} & \begin{minipage}[b]{\linewidth}\raggedright
Within-region permutation p
\end{minipage} \\
\midrule\noalign{}
\endhead
\bottomrule\noalign{}
\endlastfoot
GPT-5 & Individualizing-binding axis & 0.011 & 0.853 & 0.873 & 0.823 \\
GPT-5.4 & Individualizing-binding axis & 0.006 & 0.912 & 0.917 &
0.886 \\
Claude Opus 4.6 & Individualizing-binding axis & -0.014 & 0.765 & 0.768
& 0.908 \\
Gemini 3.1 Pro & Individualizing-binding axis & -0.008 & 0.892 & 0.904 &
0.931 \\
GPT-5 & Perceived tightness & 0.102 & 0.120 & 0.053 & 0.306 \\
GPT-5.4 & Perceived tightness & 0.058 & 0.321 & 0.188 & 0.571 \\
Claude Opus 4.6 & Perceived tightness & 0.075 & 0.165 & 0.051 & 0.466 \\
Gemini 3.1 Pro & Perceived tightness & 0.188 & 0.006 & 0.003 & 0.190 \\
\end{longtable}

\emph{Note.} Each slope comes from an ordinary least squares regression
of deviation recovery on one society-level predictor, fitted separately
for each LLM across the 88 to 89 societies with available data. The
individualizing-binding predictor is the raw mean of the nine-item
forced-choice scale; Figure S6 uses a factor score derived from the same
items. The naive p values assume independent societies, although shared
language, religion and history may induce dependence. The
region-clustered p values use a sandwich estimator with the 4 regions as
clusters and should be interpreted cautiously given the small number of
clusters. The permutation p values are based on 5,000 shuffles of the
predictor among societies within each region. These shuffles preserve
regional differences while randomizing the pairing of predictor and
outcome within regions, so the permutation test asks whether the
association is present within regions.

\needspace{8\baselineskip}
\textbf{Table S19.} Individual-level benchmark.

\begin{longtable}[]{@{}llll@{}}
\toprule\noalign{}
LLM & Mean share & Lowest society & Highest society \\
\midrule\noalign{}
\endhead
\bottomrule\noalign{}
\endlastfoot
GPT-5 & 55\% & 5\% & 94\% \\
GPT-5.4 & 46\% & 2\% & 92\% \\
Claude Opus 4.6 & 82\% & 17\% & 99\% \\
Gemini 3.1 Pro & 58\% & 4\% & 99\% \\
\end{longtable}

\emph{Note.} Each respondent is scored by the mean absolute distance
between their own ratings and the mean of the other respondents who
rated the same cell, and each LLM by the mean absolute distance between
its rating and the full cell mean. A cell rated by a single respondent
supplies no leave-one-out target, so it is dropped from the respondents'
side of the comparison, while the LLM's score uses every cell. Each
entry is the share of a society's respondents whose distance is larger
than the LLM's; the mean share averages over the 90 societies, so a
society counts once whatever its sample size. Because the reasoning
settings differ between the exploratory and the confirmatory
collections, these values should not be read as a ranking of LLM
capability.

\needspace{8\baselineskip}
\textbf{Table S20.} Within-society accuracy by region.

\begin{longtable}[]{@{}
  >{\raggedright\arraybackslash}p{(\linewidth - 12\tabcolsep) * \real{0.1618}}
  >{\raggedright\arraybackslash}p{(\linewidth - 12\tabcolsep) * \real{0.3369}}
  >{\raggedright\arraybackslash}p{(\linewidth - 12\tabcolsep) * \real{0.0587}}
  >{\raggedright\arraybackslash}p{(\linewidth - 12\tabcolsep) * \real{0.1006}}
  >{\raggedright\arraybackslash}p{(\linewidth - 12\tabcolsep) * \real{0.0546}}
  >{\raggedright\arraybackslash}p{(\linewidth - 12\tabcolsep) * \real{0.0752}}
  >{\raggedright\arraybackslash}p{(\linewidth - 12\tabcolsep) * \real{0.2122}}@{}}
\toprule\noalign{}
\begin{minipage}[b]{\linewidth}\raggedright
LLM
\end{minipage} & \begin{minipage}[b]{\linewidth}\raggedright
Measure
\end{minipage} & \begin{minipage}[b]{\linewidth}\raggedright
West
\end{minipage} & \begin{minipage}[b]{\linewidth}\raggedright
Latin America
\end{minipage} & \begin{minipage}[b]{\linewidth}\raggedright
Asia
\end{minipage} & \begin{minipage}[b]{\linewidth}\raggedright
Africa
\end{minipage} & \begin{minipage}[b]{\linewidth}\raggedright
Ordered contrast
\end{minipage} \\
\midrule\noalign{}
\endhead
\bottomrule\noalign{}
\endlastfoot
GPT-5 & Within-society r & 0.89 & 0.88 & 0.84 & 0.78 & t(88) = 9.13,
\emph{p} \textless{} .001 \\
GPT-5 & Within-society MAE, unadjusted & 0.81 & 0.87 & 0.91 & 0.99 & \\
GPT-5 & Within-society MAE, offset-adjusted & 0.70 & 0.82 & 0.78 & 0.97
& \\
GPT-5.4 & Within-society r & 0.86 & 0.86 & 0.82 & 0.76 & t(88) = 9.81,
\emph{p} \textless{} .001 \\
GPT-5.4 & Within-society MAE, unadjusted & 0.93 & 0.91 & 0.93 & 0.94
& \\
GPT-5.4 & Within-society MAE, offset-adjusted & 0.66 & 0.71 & 0.69 &
0.81 & \\
Claude Opus 4.6 & Within-society r & 0.86 & 0.87 & 0.81 & 0.76 & t(88) =
9.19, \emph{p} \textless{} .001 \\
Claude Opus 4.6 & Within-society MAE, unadjusted & 0.64 & 0.63 & 0.71 &
0.78 & \\
Claude Opus 4.6 & Within-society MAE, offset-adjusted & 0.59 & 0.60 &
0.67 & 0.78 & \\
Gemini 3.1 Pro & Within-society r & 0.91 & 0.90 & 0.86 & 0.79 & t(88) =
10.26, \emph{p} \textless{} .001 \\
Gemini 3.1 Pro & Within-society MAE, unadjusted & 0.79 & 0.83 & 0.89 &
0.94 & \\
Gemini 3.1 Pro & Within-society MAE, offset-adjusted & 0.61 & 0.73 &
0.71 & 0.88 & \\
\end{longtable}

\emph{Note.} Entries are regional means of the within-society
correlation and of two within-society MAEs: the unadjusted MAE that
Table 1 reports, and the offset-adjusted MAE (Materials and Methods)
used in the other comparisons across societies. The ordered contrast is
tested on the within-society correlation alone, in a regression across
the 90 societies (weights: West and Latin America 0.75, Asia -0.25,
Africa -1.25). A positive t means higher correlations in the West and
Latin America than in Asia and Africa. Regions are defined in
Supplementary Methods, Sample and grid.

\needspace{8\baselineskip}
\textbf{Table S21.} Agreement between the confirmatory LLMs.

\begin{longtable}[]{@{}
  >{\raggedright\arraybackslash}p{(\linewidth - 4\tabcolsep) * \real{0.4605}}
  >{\raggedright\arraybackslash}p{(\linewidth - 4\tabcolsep) * \real{0.3158}}
  >{\raggedright\arraybackslash}p{(\linewidth - 4\tabcolsep) * \real{0.2237}}@{}}
\toprule\noalign{}
\begin{minipage}[b]{\linewidth}\raggedright
LLM pair
\end{minipage} & \begin{minipage}[b]{\linewidth}\raggedright
Cross-cultural accuracy
\end{minipage} & \begin{minipage}[b]{\linewidth}\raggedright
Predicted change
\end{minipage} \\
\midrule\noalign{}
\endhead
\bottomrule\noalign{}
\endlastfoot
GPT-5.4 and Claude Opus 4.6 & 0.77 & 0.30 \\
GPT-5.4 and Gemini 3.1 Pro & 0.71 & 0.19 \\
Claude Opus 4.6 and Gemini 3.1 Pro & 0.70 & 0.49 \\
\end{longtable}

\emph{Note.} Entries are Pearson correlations between two LLMs across
scenarios. For cross-cultural accuracy, the paired values are the two
LLMs' cross-cultural correlations (defined in Table 1) for each of the
150 scenarios; for predicted change, they are their estimated changes
(Table S15) for each of the 120 scenarios with two-wave data, averaged
over societies.

\subsection{Supplementary Figures}\label{supplementary-figures}

\pandocbounded{\includegraphics[keepaspectratio]{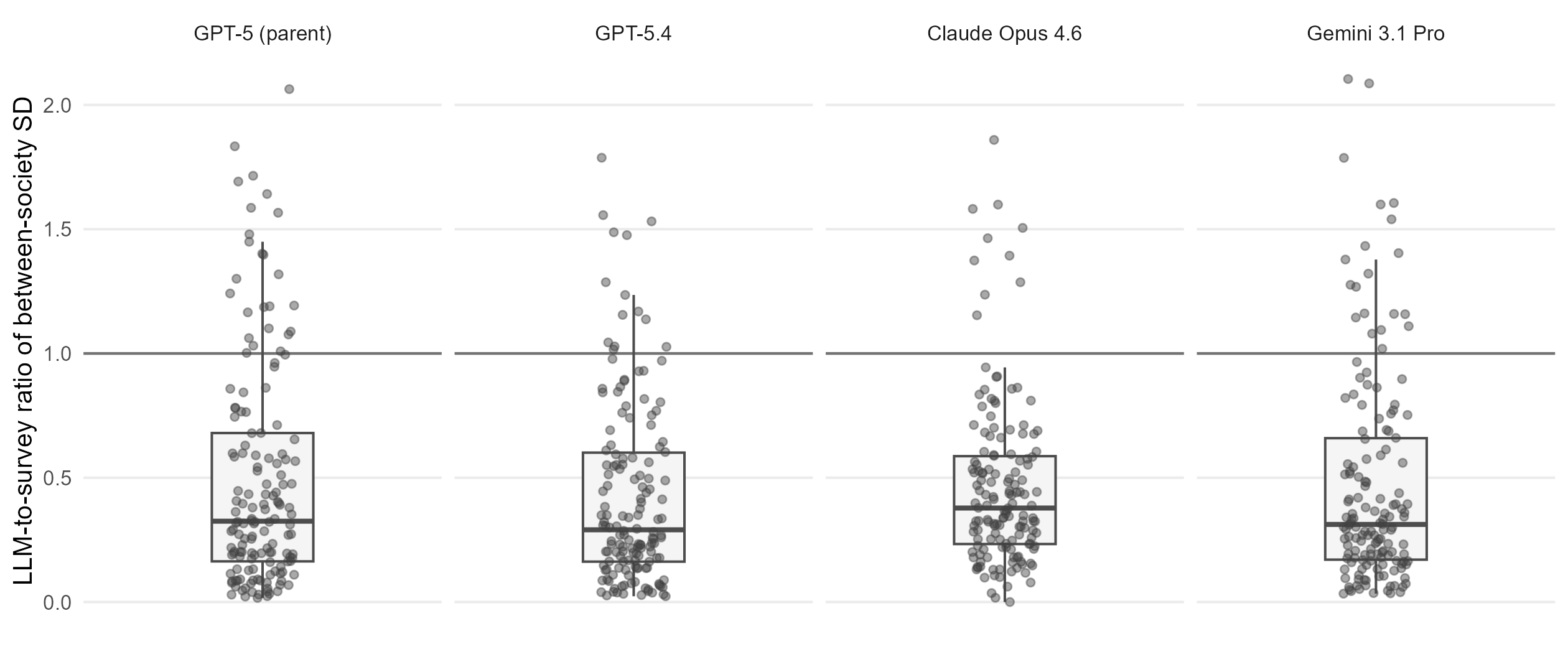}}

\textbf{Figure S1.} Between-society variation of the LLM estimates as a
share of the survey's, scenario by scenario, in each LLM: the per-model
breakdown behind the four-LLM average in Figure 1. A ratio of 1 means
equal variation, and values below 1 mean the LLM places societies too
close together. Every scenario is plotted beside its box, and no ratio
is truncated. The share of scenarios below 1 runs from 83\% to 93\%
across the four LLMs.

\pandocbounded{\includegraphics[keepaspectratio]{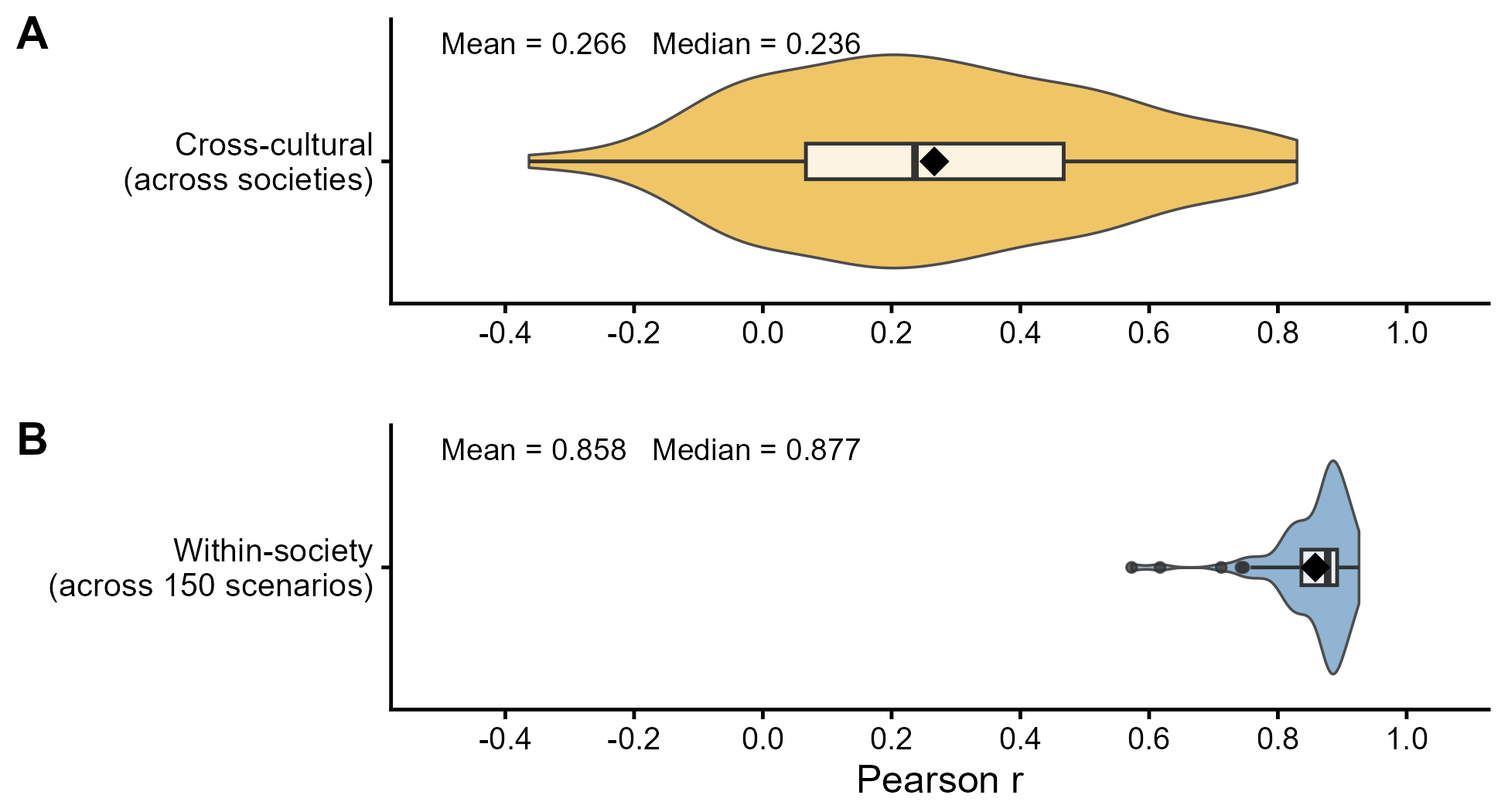}}

\textbf{Figure S2.} Distributions of cross-cultural and within-society
correlations for GPT-5. Cross-sectional benchmark on the GSEN 90 × 150
grid, shown for GPT-5 only; Table 1 gives all-LLM values. (A)
Distribution of cross-cultural correlations between GPT-5 predictions
and human ratings across the societies that rated each scenario (mean
\emph{r} = .27). (B) Distribution of within-society correlations across
150 scenarios (mean \emph{r} = .86).

\pandocbounded{\includegraphics[keepaspectratio]{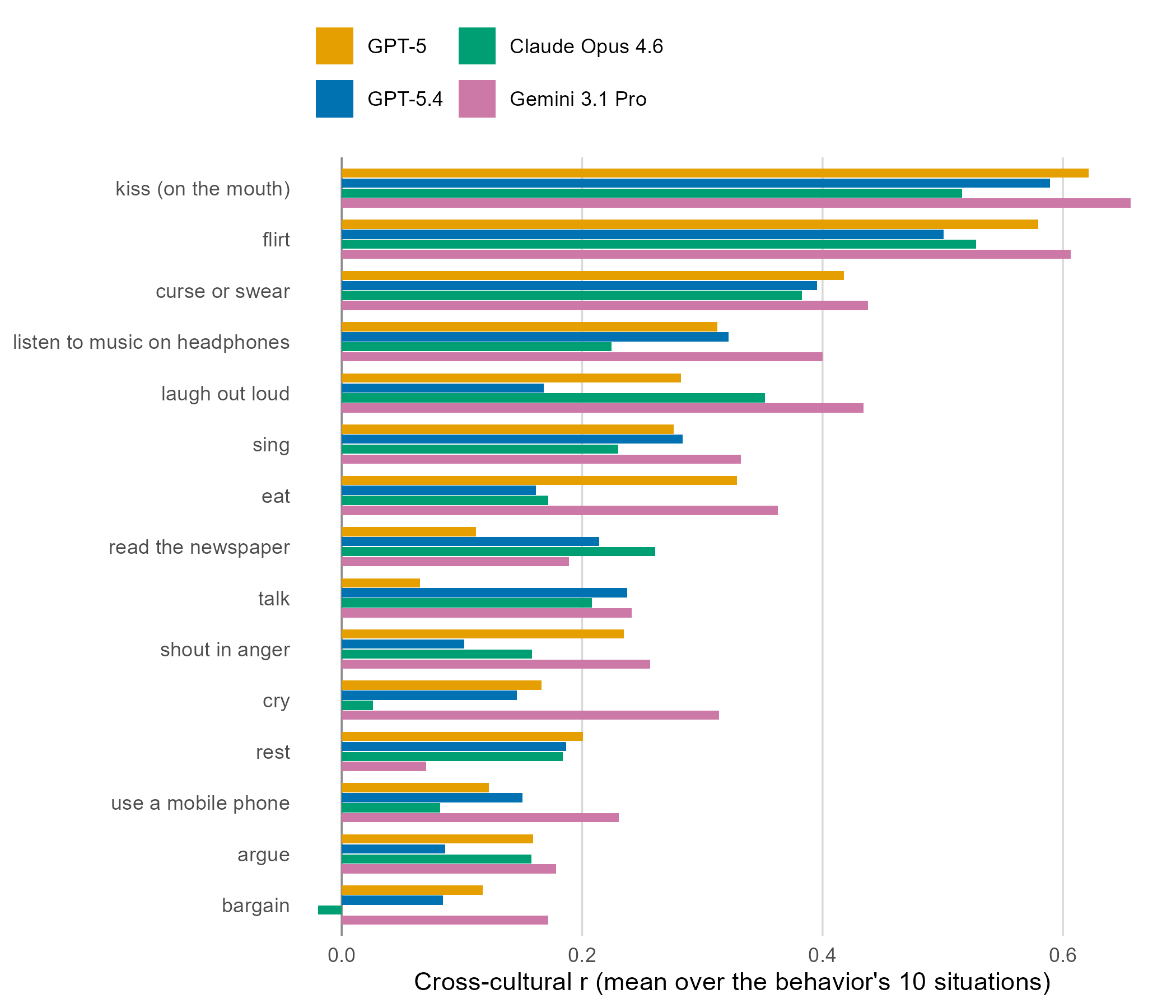}}

\textbf{Figure S3.} Cross-cultural accuracy by behavior in all four
LLMs: the per-model breakdown behind the four-LLM average in Figure 2.
Bars are the mean cross-cultural \emph{r} over the 10 situations in
which each behavior appears, and behaviors are ordered by the mean
across LLMs. Behavior means are descriptive summaries; the inferential
analyses use scenario-level observations and account for clustering
within behaviors.

\pandocbounded{\includegraphics[keepaspectratio]{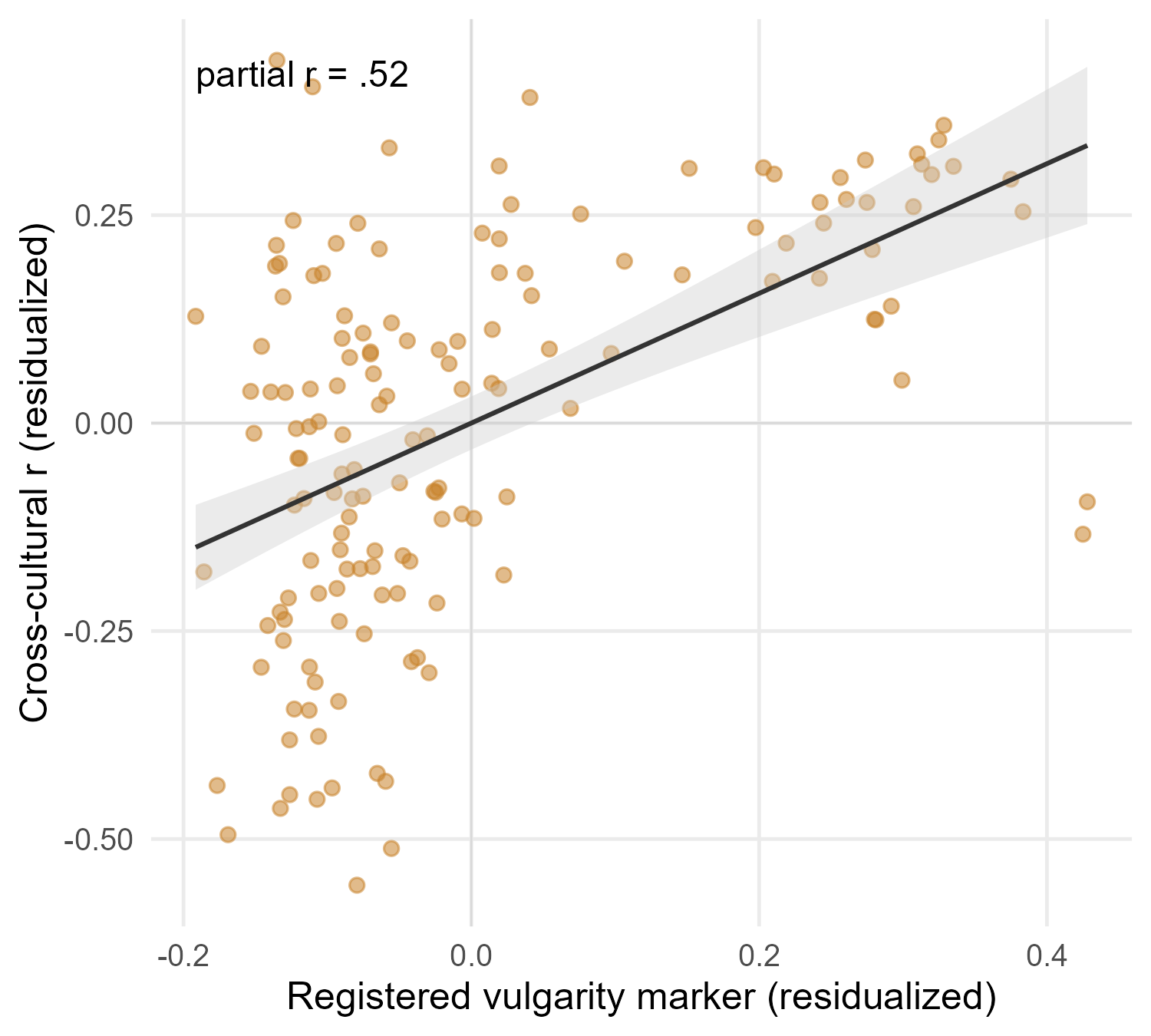}}

\textbf{Figure S4.} The vulgarity gradient in the exploratory LLM,
GPT-5, using the marker defined in the preregistration of the
confirmatory study. Added-variable plot of the rise of cross-cultural
accuracy with that marker across scenarios. Both axes give residuals
after the survey between-society SD is regressed out, so the fitted
slope is the gradient with that SD held constant.

\pandocbounded{\includegraphics[keepaspectratio]{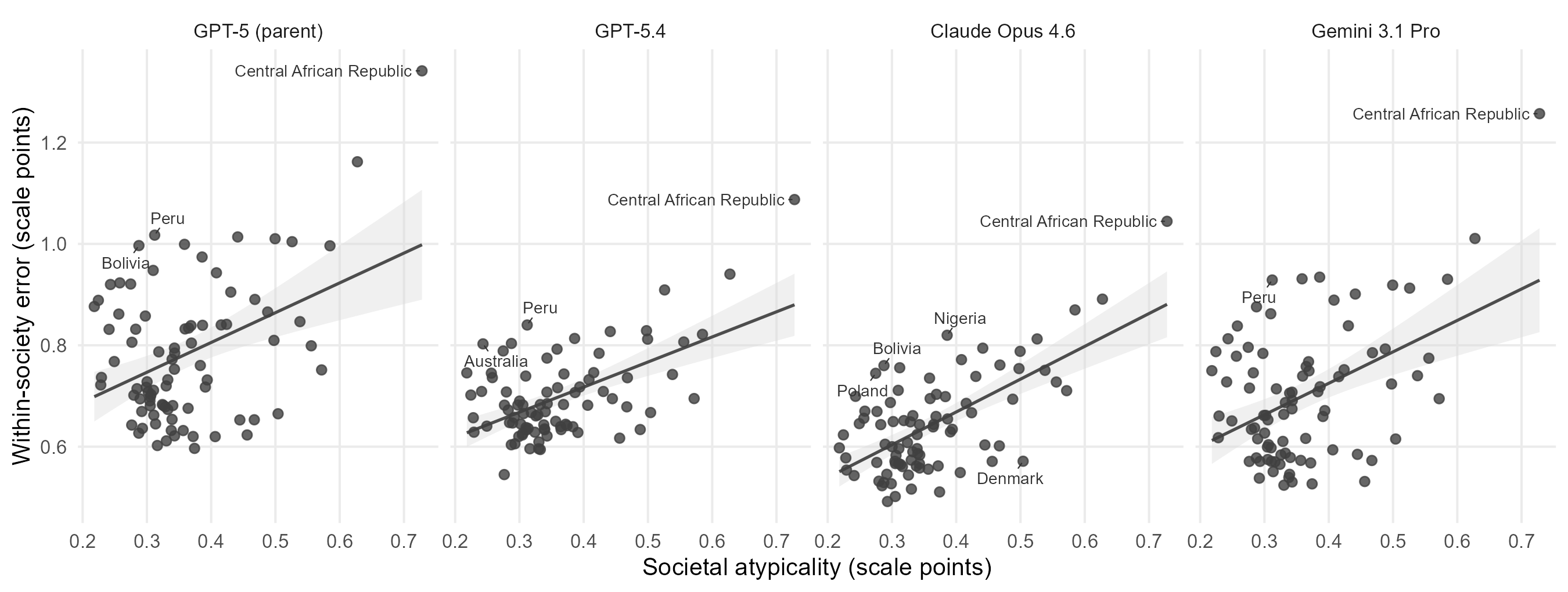}}

\textbf{Figure S5.} Within-society error against societal atypicality in
each LLM. Atypicality is the mean absolute distance between a society's
norm profile and the leave-one-out average profile of the other
societies, in points of the -2.5 to 2.5 scale. Each point is one of the
85 societies with an available Human Development Index. Both axes are
mean absolute distances taken after the society's mean signed difference
is removed: on the vertical axis between the LLM's estimates and the
survey means, on the horizontal axis between the society's own means and
the average of the other societies. Table S7 gives the correlation in
each panel. The line is an ordinary least squares fit within the panel
with its 95\% confidence band, and a society is named when its error
sits more than two residual standard deviations from that line.

\pandocbounded{\includegraphics[keepaspectratio]{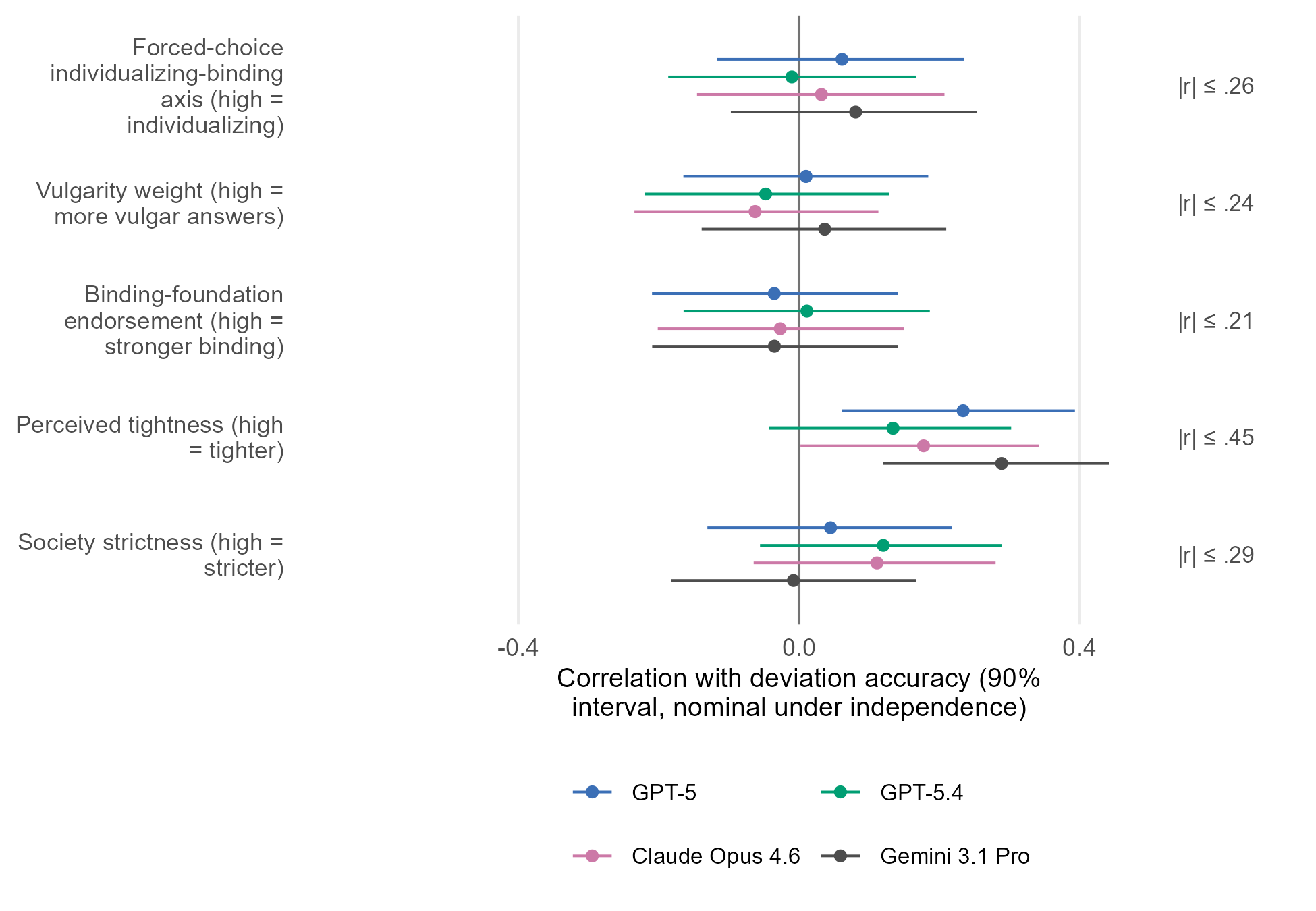}}

\textbf{Figure S6.} Correlations between deviation accuracy and five
society-level measures for each LLM, estimated across 88 to 90
societies, with nominal 90\% Fisher confidence intervals. Deviation
accuracy is the within-society correlation across scenarios between LLM
and survey deviations from their respective cross-society scenario
means. The intervals assume independent societies, although shared
language, geography and history may induce dependence. Society
strictness is computed from the same human ratings used to calculate the
outcome; the other four measures are not. The label at the right of each
row gives the largest absolute endpoint among the four LLMs' nominal
intervals, rounded outward. These labels summarize the individual
intervals and do not provide simultaneous confidence bounds across LLMs.
Table S17 reports covariate-adjusted analyses of tightness and deviation
accuracy; Table S18 tests tightness and the individualizing-binding axis
as predictors of deviation recovery under three reference distributions.

\subsection{SI References}\label{si-references}

Eriksson, K., Strimling, P., Vartanova, I., Simpson, B., Persson, M., et
al.~(2025). Everyday norms have become more permissive over time and
vary across cultures. \emph{Communications Psychology}, 3(1), 66.
https://doi.org/10.1038/s44271-025-00324-4

Gelfand, M. J., Raver, J. L., Nishii, L., Leslie, L. M., Lun, J., et
al.~(2011). Differences between tight and loose cultures: A 33-nation
study. \emph{Science}, 332(6033), 1100--1104.
\nolinkurl{https://doi.org/10.1126/science.1197754}

\end{document}